\documentclass[a4paper]{article}

\usepackage{amsthm,amsmath,amssymb,amsfonts}

\def \ring {{\cal R}}

\def\bbbn{{\mathbb N}}
\def\bbbc{{\mathbb C}}
\def\bbbz{{\mathbb Z}}

\def\diag {\mbox{diag}}

\def\hu{\hat{u}}
\def\hv{\hat{v}}

\newtheorem{Def}{Definition}
\newtheorem{The}{Theorem}
\newtheorem{Pro}{Proposition}
\newtheorem{Lem}{Lemma}
\newtheorem{Cor}{Corollary}

\begin{document}
\bibliographystyle{alpha}
\title{Symmetry structure of integrable non-evolutionary equations
\footnote{Acknowledgements. Authors would like to thank Alexander Mikhailov and Jan Sanders for
inspiring discussions and useful comments. The work of VSN was funded by the EPSRC Postdoctoral
Fellowship grant. VSN is also grateful to Vrije Universiteit Amsterdam and EU GIFT project (NEST --
Adventure Project no. 5006).}}
\author{V.S. Novikov and Jing Ping Wang\\
Institute of Mathematics and Statistics, University of Kent, UK}
\date{November 10, 2006}
 \maketitle
\begin{abstract}
 We study a class of evolutionary partial differential systems with two components
related to second order (in time) non-evolutionary equations of odd order in spatial variable. We
develop the formal diagonalisation method in symbolic representation, which enables us to derive an
explicit set of necessary conditions of existence of higher symmetries. Using these conditions we
globally classify all such homogeneous integrable systems, i.e. systems which possess a hierarchy of
infinitely many higher symmetries.

\end{abstract}

\section{Introduction.}
Classification of integrable equations of a given family of nonlinear equations is an important topic
in the field of soliton theory. There are many approaches to this problem, among which the symmetry
approach has proved to be very efficient and powerful method. The symmetry approach has been used to
classify large classes of integrable nonlinear partial differential equations and
difference-differential equations such as scalar evolutionary equations, Volterra and Toda type
equations, hyperbolic equations etc. We refer the reader to the recent review paper \cite{asy} and the
references therein for details.

This paper is devoted to a problem of classification of systems of
integrable evolutionary equations with two components. The first
work in this direction had been carried out by Mikhailov, Shabat and
Yamilov \cite{ms1,ms2,msy}. They classified integrable second order
systems of the form
$$ {\bf u}_t={\bf A}({\bf u}) {\bf u}_{xx}+F({\bf u},{\bf u}_x),
\quad \det({ A}({\bf u}))\neq 0, \quad {\bf u}=(u,\ v)^T ,
$$
where the $2\times 2$ matrix $A({\bf u})$   can be reduced to
$${ A}({\bf u})=\mbox{diag}(a, \ b), \quad a=-1, \quad b=1,$$
if the system possesses higher order conservation laws. Later, Svinolupov \cite{Svi} studied the
Burgers type equations, i.e. the case \(a=b=1\) (see also the  review paper \cite{mss}). After these
two special cases, attention was turned to the more general case: $a\neq \pm b$ and $ab\neq 0$
\cite{bp} and to the systems of higher order. In principle, the symmetry approach can be applied to
classification of higher order systems. Although the method is algorithmic, the computation involved
grows very rapidly with the order of the system, so further achievement in this direction had to wait
for the development of the appropriate computer algebra.

Recently, several research groups have developed the computer programmes to carry out the task of
classification of polynomial integrable evolutionary systems with two components. Foursov and Olver
considered the symmetrically-coupled evolutionary equations of lower order \cite{fo}. After linear
transformation, the leading linear part of the systems can always be diagonalised.  Tsuchida and Wolf
studied polynomial homogeneous integrable evolutionary systems of mixed scalar and vector dependent
variables \cite{tw} of order $2$ and $3$, where the authors compared their comprehensive results with
the existing ones. The linear parts of systems they considered are diagonal including zero eigenvalues.
Their classification methods have led to interesting new integrable equations, but they could not claim
that their lists are complete in the sense that there are no other integrable systems of the fixed
order in certain classes since the methods they used were based on searching symmetries of specific
orders and forms. The symbolic representation combined  with applications of number theory \cite{jp} is
the solution to this problem at least for subclass of homogeneous polynomial evolutionary systems. It
enables us to translate the question of integrability into the problem of divisibility of certain
symmetric polynomials, which can be solved by using number theoretical methods. A series papers have
been published on it including the complete classification of second order two components evolutionary
systems \cite{sw} and Bakirov-type systems \cite{fsw,bsw}.

The ultimate goal of classification of integrable equations  is to
obtain the classification result for integrable equations of any
order, i.e. the global classification result. So far the only global
result has been obtained for scalar homogeneous evolutionary
equations \cite{sw1,sw2}. In this paper, based on the symbolic
representation in \cite{mnw1}, we develop the formal diagonalisation
method so that we can globally classify homogeneous two-component
evolutionary systems whose leading linear terms are not explicitly
diagonal. We demonstrate our method  classifying integrable systems
of the form
\begin{eqnarray}
\left\{\begin{array}{l}\label{eq1} u_t=\partial_x^r v,\\
v_t=\partial_x^{2n+1-r}u+F(u,u_x,u_{xx},...,\partial^{2n-r}_x
u,v,v_x,v_{xx},...,\partial^{2n-r}_x v),\quad n=1,2,3,\ldots,\quad
r\in \{0,1,\ldots,n\},
\end{array}\right.
\end{eqnarray}
of arbitrary order (i.e. for any integer $n>0$). Here $F$ is a
homogeneous differential polynomial in $u,v$ and their
$x$-derivatives. If function $F$ does not depend on
$v,v_{x},\ldots,\partial_x^{r-1}v$ then we can exclude $v$ from
system (\ref{eq1}) and obtain non-evolutionary equations of odd
order
\begin{equation}
\label{eq1e}
u_{tt}=\partial_x^{2n+1}u+K(u,u_x,u_{xx},...,\partial^{2n}_x
u,u_t,u_{t\, x},u_{t\, xx},...,\partial^{2n}_x u_{t}),\qquad
K=D_x^r(F),\qquad n=1,2,3,\ldots
\end{equation}

The even-order non-evolutionary equations of form (\ref{eq1e}) have
been studied using the perturbative symmetry approach in symbolic
representation in \cite{mnw1}. The famous Boussinesq equation
\cite{Zak}
\[
u_{tt}=u_{xxxx}+(u^2)_{xx}
\]
belongs to this class. As a result,  integrable equations of orders $4$ and $6$ were classified as well
as three new integrable equations of order $10$ were found. The approach developed in \cite{mnw1} is
also suitable for odd order equations. However, there is a difficulty to formulate the explicit
necessary conditions for this case. Besides, it can not be used for global classification. The special
subclass of non-evolutionary equations was also considered in \cite{mnw}, where a family of partially
integrable equations with only one higher symmetry was constructed.

In this paper in the framework of the perturbative symmetry approach in the symbolic representation we
derive an explicit set of necessary conditions of existence of a hierarchy of higher symmetries for
systems (\ref{eq1}). These conditions are obtained after diagonalisation of  the linear part of system
(\ref{eq1}) via a non-local linear transformation of dependent variables $u,v$. We prove the ultimate
classification result: If a homogeneous system (\ref{eq1}) with positive weight of $u$ possesses a
hierarchy of infinitely many higher symmetries, then up to re-scaling it one of the following two
equations:
\begin{eqnarray*}
&&\left\{\begin{array}{l} u_t=v_x,\\
v_t=u_{xx}+3uv_x+vu_x-3u^2u_x,
\end{array}\right.\\
&&\left\{\begin{array}{l} u_t=v_x,\\
v_t=(\partial_x+u)^{2n}(u)-v^2,\quad n=1,2,3,\ldots .
\end{array}\right.
\end{eqnarray*}
These systems can be rewritten in the form of non-evolutionary
equations if we introduce a new variable $u=w_x$:
\[
w_{tt}=w_{xxx}+3w_xw_{t,x}+w_{xx}w_t-3w_x^2w_{xx}.
\]
\[
w_{tt}=(\partial_x+w_x)^{2n}(w_x)-w_t^2
\]
The first equation is known to be integrable \cite{hss}. The second
equation can be brought into linear $f_{tt}=\partial_x^{2n+1} f$ by
the Cole-Hopf transformation $w=\log(f)$.

The paper is organized as follows: in section \ref{Sec2} we remind
basic definitions and notations of the symbolic representation
\cite{mnw1}; in section \ref{Sec3} we develop the formal
diagonalisation formalism for systems (\ref{eq1}); using the formal
diagonalisation method we obtain explicit formulae for the
(approximate) symmetries of system (\ref{eq1}) and derive necessary
conditions for the existence of an infinite hierarchy of higher
symmetries; in section \ref{Sec4} we apply these conditions to
classify integrable homogeneous systems (\ref{eq1}) for all orders
$n=1,2,3,\ldots$ and for all possible representations
$r=0,1,\ldots,n$.

\section{Symbolic representation over the ring of differential polynomials.}
\label{Sec2} In this section we remind basic definitions and
notations of the ring of differential polynomials and symbolic
representation (for more details see e.g. \cite{sw, mnw1, mn}).

Throughout the whole paper we assume that all functions, such as $F$ and $K$ in equation (\ref{eq1})
and (\ref{eq1e}), are differential polynomials in variables $u,v,u_x,v_x,u_{xx}, \ldots$ . We introduce
an infinite sequence of {\em dynamical variables} $\{ u_0,v_0,u_1,v_1,u_2,v_2,\cdots \}$ by the
identification
\begin{equation}\label{dynvars}
 u_0=u,\ v_0=v,\ u_n=\partial^n_x u,\ v_n=\partial^n_x v.
\end{equation}
We often omit the zero index and write $u$ and $v$ instead of $u_0$
and $v_0$.

We denote $\ring$ the ring of polynomials over $\bbbc$ of infinite
number of dynamical variables. We also  assume that $1\not\in
\ring$. Elements of the ring $\ring$ are finite sums of monomials
with complex coefficients and therefore each element depends on a
finite number of the dynamical variables. The degree of a monomial
is defined as a total power, i.e. the sum of all powers of dynamical
variables that contribute to the monomial. Let ${\cal R}^n$ denote
the set of polynomials in $\ring$ of degree $n$. The ring $\ring$
has a natural gradation
\[ \ring = \bigoplus_{n\in \bbbz_+}{\cal R}^n \, ,
\quad {\cal R}^n \cdot {\cal R}^m\subset {\cal R}^{n+m} \, .
\]
 Elements of ${\cal R}^1$ are linear
functions of the dynamical variables, ${\cal R}^2$ quadratic, etc.
We suppose that $1\notin\ring$. It is convenient to define a
``little-oh'' order symbol $o({\cal R}^n)$. We say that $f=o({\cal
R}^n)$ if $f\in \bigoplus_{k>n}{\cal R}^k$, i.e. the degree of every
monomial of $f$ is bigger than $n$.

Let $\mu, \nu$ be two positive rational numbers\footnote{In principle, the weights could be any
rational numbers including zero, but in this paper we consider only positive weights.}, which we call
the weights of $u$ and $v$ respectively and denote $W(u)=\mu, W(v)=\nu$. We define the weights of
dynamical variables (\ref{dynvars}) as $W(u_i)=\mu+i$ and $W(v_j)=\nu+j$. The weight of a monomial is
the sum of the weights of dynamical variables that contribute to the monomial. We say that a polynomial
$f\in\ring$ is a homogeneous polynomial of weight $\lambda$  (and write $W(f)=\lambda$) if every its
monomial is of weight $\lambda$. The {\sl weighted} gradation of $\ring$ is defined as
\[
\ring = \bigoplus_{m,n,s\in \bbbz_{\ge 0}}{\cal R}_{m\mu+n\nu+s} \,
, \quad\quad {\cal R}_p \cdot {\cal R}_q\subset {\cal R}_{p+q} \, ,
\]
where ${\cal R}_p$ is the set of polynomials in $\ring$ of weight $p$.
In fact, we can define a degree-weighted
gradation of the ring $\ring$
\[ \ring = \bigoplus_{n,m,s\in \bbbz_{\ge 0},\,r\in\bbbz_+}{\cal R}_{m\mu+n\nu+s}^r \, ,
\quad {\cal R}_p^n \cdot {\cal R}_q^m\subset {\cal R}_{p+q}^{n+m} \, ,
\]
where ${\cal R}_p^r$ is the set of polynomials in $\ring$ of degree
$r$ and weight $p$. Note that the condition $\mu,\nu>0$ makes each
such subspace to be finitely generated. For example, if $\mu=1$,
$\nu=2$ then
\[
{\cal R}^2_4=\mbox{span}\{uu_2,u_1^2,u_1v,uv_1,v^2\}
\]

The ring $\ring$ is a differential ring with a derivation defined as
\begin{equation}\label{Dx}
 D_x =\sum _{k\ge 0} \left( u_{k+1} \frac{\partial}{\partial u_{k}}
 +v_{k+1}\frac{\partial}{\partial v_{k}}\right) \, .
\end{equation}
Since $1\not\in\ring$, the kernel of the linear map
$D_x:\ring\mapsto\mbox{Im}\, D_x\subset\ring$ is empty and
therefore $D_x^{-1}$ is defined uniquely on $\mbox{Im}\, D_x$. It is easy to verify that
\[ D_x :{\cal R}_p^n\longmapsto {\cal R}_{p+1}^n\, .\]

To any element $g\in\ring$ we associate  differential operators $g_{*,u}$ and $g_{*,v}$ called
Fr\'echet derivatives with respect to $u$ and $v$ and  defined as
\[
g_{*,u}=\sum_{k\ge 0}\frac{\partial g}{\partial u_k}D_x^k,\qquad
g_{*,v}=\sum_{k\ge 0}\frac{\partial g}{\partial v_k}D_x^k.
\]

Now we define the symbolic representation $\hat{\cal R}$ of the ring $\ring$. A symbolic representation
of a monomial $$u_0^{n_0}u_1^{n_1}\cdots u_p^{n_p}v_0^{m_0}v_1^{m_1}\cdots v_q^{m_q},\, n_0+n_1+\cdots
+n_p=n,\, m_0+m_1+\cdots+m_q=m$$ is defined as:
\begin{eqnarray}\nonumber
 &&u_0^{n_0}u_1^{n_1}\cdots
u_p^{n_p}v_0^{m_0}v_1^{m_1}\cdots v_q^{m_q}\to\\ \label{mon} &&\to
u^nv^m
\langle\xi_1^0\xi_2^0\cdots\xi_{n_0}^0\xi_{n_0+1}^1\cdots
\xi_{n_0+n_1}^1\cdots\xi_n^p\rangle_{\xi}\langle\zeta_1^0\zeta_2^0\cdots
\zeta_{m_0}^0\zeta_{m_0+1}^1
\cdots\zeta_{m_0+m_1}^1\cdots\zeta_m^q\rangle_{\zeta}\ ,
\end{eqnarray}
where triangular brackets $\langle\rangle_{\xi}$ and
$\langle\rangle_{\zeta}$ denote the averaging over group $\Sigma_n$
of permutations of $n$ elements $\xi_1,\ldots,\xi_n,$ , i.e.
\[
\langle c(\xi_1,\ldots,\xi_n,\zeta_1,\ldots
,\zeta_m)\rangle_{\xi}=\frac{1}{n!}\sum_{\sigma\in\Sigma_n}c(\sigma(\xi_1),\ldots,\sigma(\xi_n),\zeta_1,\ldots
,\zeta_m)
\]
and group $\Sigma_m$ of $m$ elements $\zeta_1,\ldots,\zeta_m$
respectively. Later we refer to this as symmetrisation operation.
For example, linear monomials $u_n, v_m$ are represented by
\begin{equation}\label{unvm}
u_n\to u\xi_1^n,\quad v_m\to v\zeta_1^m
\end{equation}
and quadratic monomials $u_nu_m$, $u_nv_m$, $v_nv_m$  have the following symbols
\begin{equation}\label{monom}
u_nu_m\to\frac{u^2}{2}(\xi_1^n\xi_2^m+\xi_1^m\xi_2^n),\quad u_nv_m\to uv(\xi_1^n\zeta_1^m),
\quad
v_nv_m\to \frac{v^2}{2} (\zeta_1^n\zeta_2^m+\zeta_1^m\zeta_2^n)\ .
\end{equation}
To the sum of two elements of the ring corresponds  the sum of their
symbols. To the product of two elements $f,g\in\ring$ with symbols
$f\to u^nv^ma(\xi_1,\ldots,\xi_n,\zeta_1,\ldots,\zeta_m)$ and $g\to
u^pv^qb(\xi_1,\ldots,\xi_p,\zeta_1,\ldots,\zeta_q)$ corresponds:
\begin{equation}\label{multmonoms}
fg\to u^{n+p}v^{m+q}\langle\langle
a(\xi_1,\ldots,\xi_n,\zeta_1,\ldots,\zeta_m)b(\xi_{n+1},\ldots,\xi_{n+p},
\zeta_{m+1},\ldots,\zeta_{m+q})\rangle_{\xi}\rangle_{\zeta},
\end{equation} where the symmetrisation operation is taken with respect to
permutations of all arguments $\xi$ and $\zeta$. It is easy to see
that the symbolic representations of quadratic (\ref{monom}) and general
(\ref{mon}) monomials immediately follows from (\ref{unvm}) and
(\ref{multmonoms}).

If $f\in\ring$ has a symbol $f\to
u^nv^ma(\xi_1,\ldots,\xi_n,\zeta_1,\ldots,\zeta_m)$, then the
symbolic representation for its $N$-th derivative $D_x^N(f)$ is:
\[
D_x^N(f)\to
u^nv^m(\xi_1+\xi_2+\cdots+\xi_n+\zeta_1+\zeta_2+\cdots\zeta_m)^N
a(\xi_1,\ldots,\xi_n,\zeta_1,\ldots,\zeta_m) .
\]
We will assign a symbol $\eta$ to the operator $D_x$ in the symbolic
representation with obvious action rule
\[
\eta^N(u^nv^ma(\xi_1,\ldots,\xi_n,\zeta_1,\ldots,\zeta_m))=u^nv^m
(\xi_1+\xi_2+\cdots+\xi_n+\zeta_1+\zeta_2+\cdots\zeta_m)^N
a(\xi_1,\ldots,\xi_n,\zeta_1,\ldots,\zeta_m)
\]
Note that $N$ in the above formula can be not only positive integer but in principal any rational
number (or even a complex number) as long as such calculation has any sense apart from formal . In the
next section we will need $N$ to be half-integer numbers which correspond to fractional derivation in
$x$-space.

If $g\in\ring$ and $g\to u^nv^m
a_{n,m}(\xi_1,\ldots,\xi_n,\zeta_1,\ldots,\zeta_m)$ then for the
symbol of its Fr{\'e}chet derivatives $g_{*,u}$ and $g_{*,v}$ we
have
\[
g_{*,u}\to
nu^{n-1}v^ma_{n,m}(\xi_1,\ldots,\xi_{n-1},\eta,\zeta_1,\ldots,\zeta_m),\qquad
g_{*,v}\to
mu^nv^{m-1}a_{n,m}(\xi_1,\ldots,\xi_n,\zeta_1,\ldots,\zeta_{m-1},\eta).
\]

Thus we obtained the symbolic representation $\hat{\cal R}$ of the
differential ring $\ring$.

\section{Structure of symmetries and approximate symmetries.}\label{Sec3}
Consider a family of homogeneous polynomial systems (\ref{eq1}), i.e.
\begin{eqnarray}
\left\{\begin{array}{l}\label{eq2} u_t=\partial_x^r v,\\
v_t=\partial_x^{2n+1-r}u+F(u,u_1,\ldots,u_{2n-r},v,v_1,\ldots,v_{2n-r}),\quad
n=1,2,3,\ldots,\quad r\in \{0,1,\ldots,n\}.
\end{array}\right.
\end{eqnarray}

\begin{Def}\label{Def1}
A pair of differential polynomials $G$ and $ M$is called a symmetry of an evolution system
$u_{t}=f,\quad v_{t}=g,$ where $f,\ g,\ G$ and $M$ are all in ring $\ring$, if system $
u_{\tau}=G,\quad v_{\tau}=M $ is compatible with the given system.
\end{Def}

This definition is equivalent to the Lie bracket between ${\bf a}= (f,g)^{\mbox{tr}}$ and ${\bf
b}=(G,M)^{\mbox{tr}}$ vanishing, where the bracket is defined as
\begin{eqnarray}\label{lieD}
 [{\bf a},{\bf b}]=\left(\begin{array}{c}
f_{*,u}(G)+f_{*,v}(M)-G_{*,u}(f)-G_{*,v}(g)\\
g_{*,u}(G)+g_{*,v}(M)-M_{*,u}(f)-M_{*,v}(g)
\end{array}\right)\, .
\end{eqnarray}
The compatibility condition for system (\ref{eq2}) can be written as
\begin{eqnarray}\label{symd}
D_t(G)=D_x^r v_{\tau}=D_x^r M, \quad D_t(M)=D_{\tau}(u_{2n+1-r}+F),
\end{eqnarray}
where evolutionary derivations $D_t$ and $D_{\tau}$ are
\[
D_t=\sum_{j\ge 0}\left(v_{j+r}\frac{\partial}{\partial u_j}+D_x^j(u_{2n+1-r}+F)
\frac{\partial}{\partial
v_j}\right),\quad D_{\tau}=\sum_{j\ge 0}\left(D_x^j(G)\frac{\partial}{\partial
u_j}+D_x^{j-r}D_t(G)\frac{\partial}{\partial v_j}\right)\ .
\]
Eliminating $M$ from these equations, we obtain
\begin{eqnarray*}
D_x^{-r}D_t^2(G) =D_{\tau}(u_{2n+1-r}+F),
\end{eqnarray*}
Thus the symmetry of system (\ref{eq2}) is completely determined by its first component $G\in\ring$.

\begin{Def}\label{Def2}
We say that a differential polynomial $G\in\ring$ generates an
approximate symmetry of degree $p$ of system (\ref{eq2}) if $G$
satisfies the equation:
\[
D_x^{-r}D_t^2(G)-D_{\tau}(u_{2n+1-r}+F)=o(\ring^p).
\]
\end{Def}

Notice that any system (\ref{eq2}) has an infinite hierarchy of
approximate symmetries of degree $1$. These are simply the
symmetries of its linear part $u_t=v_r,\,\,v_t=u_{2n+1-r}$. The
requirement of the existence of approximate symmetries of degree $2$
is very restrictive. Integrable equations by definition possess
infinite hierarchies of approximate symmetries of any degree. We
will prove later in this paper that the existence of infinite
hierarchy of approximate symmetries of degree $2$ and the existence
of at least one exact symmetry for systems (\ref{eq2}) implies
integrability just as in the case of scalar evolutionary equations
\cite{sw}.

We now derive the necessary and sufficient conditions of the
existence of approximate symmetries of degree $2$. It is convenient
to do this in the symbolic representation.

System (\ref{eq2}) in the symbolic representation takes the form
\begin{eqnarray}
\left\{\begin{array}{l}\label{eqs} u_t=v\zeta_1^r,\\
v_t=u\xi_1^{2n+1-r}+\sum_{k\ge 2}\sum_{i=0}^k
u^iv^{k-i}a_{i,k-i}(\xi_1,\ldots,\xi_i,\zeta_1,\ldots,\zeta_{k-i}) .
\end{array}\right.
\end{eqnarray}
Let us assume that the weight of variable $u$, denoted by $w$, is
positive, i.e. $W(u)=w>0$. Since the system is homogeneous, we have
\begin{eqnarray*}
\left\{\begin{array}{l} w+W(D_t)=W(v)+r\\
W(v)+W(D_t)=2n+1-r+w,\end{array}\right.
\end{eqnarray*}
which leads to $W(D_t)=n+\frac{1}{2}$ and $W(v)=w+n+\frac{1}{2}-r$.

Homogeneous polynomials
$a_{i,j}(\xi_1,\ldots,\xi_i,\zeta_1,\ldots,\zeta_j)$ are symmetric
with respect to variables $\xi_1,\ldots,\xi_i$ and
$\zeta_1,\ldots,\zeta_j$. The degrees of $a_{i,j}$ are given by
\begin{equation}
\label{deg} \deg(a_{i,j})=2n+1-r+w-i W(u)-jW(v)=(2-j)n-(i+j-1)w+(j-1)r+\frac{2-j}{2}.
\end{equation}
The degree of $a_{i,j}$ determined above must be a  non-negative
integer, otherwise, we put $a_{i,j}=0$. The sum in (\ref{eqs})
terminates due to the assumption $W(u)=w>0$.

The symmetry of system (\ref{eq2}) always starts with linear terms. A homogeneous symmetry of
(\ref{eq2}) in the symbolic representation starts either with $u\xi_1^m$ or with $v\zeta_1^{m+r}$.
Without loss of generality we have
\begin{equation}
\label{sym1} u_{\tau}=G=u\xi_1^m+\sum_{s\ge 2}\sum_{j=0}^s
u^jv^{s-j}A_{j,s-j}(\xi_1,\ldots,\xi_j,\zeta_1,\ldots,\zeta_{s-j}),\quad
m>1
\end{equation}
or
\begin{equation}
\label{sym2} u_{\tau}=G=v\zeta_1^{m+r}+\sum_{s\ge 2}\sum_{j=0}^s
u^jv^{s-j}A_{j,s-j}(\xi_1,\ldots,\xi_j,\zeta_1,\ldots,\zeta_{s-j}),\quad
m>0
\end{equation}
We call the symmetries of the form (\ref{sym1}) as {\bf type I}
symmetries and those of the form (\ref{sym2}) -- {\bf type II}
symmetries. Without causing confusion, we call integer $m$ in
(\ref{sym1}) or (\ref{sym2}) the order of the corresponding
symmetry. Functions
$A_{i,j}(\xi_1,\ldots,\xi_i,\zeta_1,\ldots,\zeta_j)$ in (\ref{sym1})
and (\ref{sym2}) are homogeneous polynomials in their variables,
symmetric with respect to arguments $\xi_1,\ldots,\xi_i$ and
$\zeta_1,\ldots,\zeta_j$. These functions can be explicitly
determined in the terms of system (\ref{eqs}) from the compatibility
conditions.

\subsection{Formal diagonalisation.}
Let us first concentrate on how to compute the Lie bracket defined as (\ref{lieD}) between the linear
part of system (\ref{eqs}) denoted by $K^1$, i.e.
\begin{eqnarray}
K^1=\left(\begin{array}{c}v\zeta_1^r\\u\xi_1^{2n+1-r}\end{array}\right)
=L\left(\begin{array}{c}u\\v\end{array}\right), \quad\quad
L=\left(\begin{array}{cc} 0& \eta^r\\ \eta^{2n+1-r}&
0\end{array}\right)
\end{eqnarray}
and any pair of differential polynomials. We know its symbolic representation takes simple and elegant
form if matrix $L$ is diagonal \cite{sw}. Inspired by this, we shall formally diagonalise matrix $L$,
produce the required formula in new variables and then transform back to the original variables.

Notice that matrix $L$ has two eigenvalues
$\pm\eta^{n+\frac{1}{2}}$. Therefore, there exists a linear
transformation
\[
T=\left(\begin{array}{cc}1&1\\ \eta^{n+\frac{1}{2}-r}&
-\eta^{n+\frac{1}{2}-r}\end{array}\right)\]
such that
\[
T^{-1}LT=\diag(\eta^{n+\frac{1}{2}},-\eta^{n+\frac{1}{2}}).
\]
Let us introduce new variables $\hu$ and $\hv$
\[
\left(\begin{array}{c}u\\v\end{array}\right)=T\left(\begin{array}{c}\hu\\\hv\end{array}
\right)=\left(\begin{array}{c}
\hu+\hv\\
\hu\xi_1^{n+\frac{1}{2}-r}-\hv\zeta_1^{n+\frac{1}{2}-r}\end{array}\right).
\]
Equally, we have
\[
\left(\begin{array}{c}\hu\\\hv\end{array}\right)=T^{-1}\left(\begin{array}{c}u\\v
\end{array}\right)=\frac{1}{2}
\left(\begin{array}{c}u+\eta^{-n-\frac{1}{2}+r}(v)\\
u-\eta^{-n-\frac{1}{2}+r}(v)\end{array}\right).
\]
The new variables $\hu$ and $\hv$ have the same weights, i.e. $W(\hu)=W(\hv)=W(u)$. Without causing
confusion we assign the same symbols $\xi$ and $\zeta$ for the symbolic representation of the ring
generated by $\hu$, $\hv$ and their derivatives. The exponents of symbols can be half-integer, which
corresponds to half-differentiation in $x$-space.

Before we work out the exact form of system (\ref{eqs}) in variables
$\hu$ and $\hv$, we prove two useful propositions.

\begin{Pro}\label{Pro1}
Under the transformation $T$, an arbitrary polynomial
$u^iv^jb_{i,j}(\xi_1,\ldots,\xi_i,\zeta_1,\ldots,\zeta_j)$ takes the
form
\begin{eqnarray}
\label{termtr}
&&u^iv^jb_{i,j}(\xi_1,\ldots,\xi_i,\zeta_1,\ldots,\zeta_j)=\sum_{p=0}^i
\sum_{q=0}^j\hu^{p+q}\hv^{i+j-p-q}
C_i^pC_j^q(-1)^{j-q}\\ \nonumber && \langle\langle
b_{i,j}(\xi_1,\ldots,\xi_p,\zeta_1,\ldots,\zeta_{i-p},\xi_{p+1},\ldots,\xi_{p+q},
\zeta_{i-p+1},\ldots,\zeta_{i+j-p-q})
(\xi_{p+1}\cdots\xi_{p+q}\zeta_{i-p+1}\cdots\zeta_{i+j-p-q})^{n+\frac{1}{2}-r}
\rangle_{\xi}\rangle_{\zeta},
\end{eqnarray}
where $C_i^j$ are binomial coefficients defined by $
C_i^j=\frac{i!}{j! (i-j)!}. $
\end{Pro}
{\bf Proof}. We prove the statement by induction on both $i$ and
$j$. It is easy to see that linear terms $u\xi_1^i,\,v\zeta_1^j$
transform as
\[
u\xi_1^i=\eta^i(\hu+\hv)=\hu\xi_1^i+\hv\zeta_1^i,\quad\quad
v\zeta_1^j=\eta^j(\hu\xi_1^{n+\frac{1}{2}-r}-\hv\zeta_1^{n+\frac{1}{2}-r})=
\hu\xi_1^{n+\frac{1}{2}-r+j}-\hv\zeta_1^{n+\frac{1}{2}-r+j}.
\]
The procedure is: we first substitute $\xi_1$ and $\zeta_1$ by $\eta$, the symbol of the total
$x$-derivative $D_x$,  $u$ and $v$ by $\hu+\hv$ and
$\hu\xi_1^{n+\frac{1}{2}-r}-\hv\zeta_1^{n+\frac{1}{2}-r}$; and then we compute the action
of $\eta$ on
$\hu+\hv$ and $\hu\xi_1^{n+\frac{1}{2}-r}-\hv\zeta_1^{n+\frac{1}{2}-r}$. For the higher degree
terms,
we apply this procedure for every element in the arguments. Assume formula (\ref{termtr}) without
symmetrisation is true for arbitrary $i$ and $j$. We have
\begin{eqnarray*}
&&u^{i+1}v^jb_{i+1,j}(\xi_1,\ldots,\xi_i,\xi_{i+1},\zeta_1,\ldots,\zeta_j)\\
&=&\sum_{p=0}^i\sum_{q=0}^j\hu^{p+q}\hv^{i+j-p-q}
C_i^pC_j^q(-1)^{j-q}(\xi_{p+1}\cdots\xi_{p+q}\zeta_{i-p+1}\cdots\zeta_{i+j-p-q})^{n+\frac{1}{2}-r}\\
&&
b_{i+1,j}(\xi_1,\ldots,\xi_p,\zeta_1,\ldots,\zeta_{i-p},\eta,\xi_{p+1},\ldots,\xi_{p+q},
\zeta_{i-p+1},\ldots,\zeta_{i+j-p-q}) (\hu+\hv)
\\&=&
\sum_{p=0}^i\sum_{q=0}^j\hu^{p+q+1}\hv^{i+j-p-q}
C_i^pC_j^q(-1)^{j-q}(\xi_{p+2}\cdots\xi_{p+q+1}\zeta_{i-p+1}\cdots\zeta_{i+j-p-q})^{n+\frac{1}{2}-r}\\
&&
b_{i+1,j}(\xi_1,\ldots,\xi_p,\zeta_1,\ldots,\zeta_{i-p},\xi_{p+1},\ldots,\xi_{p+q+1},\zeta_{i-p+1},
\ldots,\zeta_{i+j-p-q})
\\
&+& \sum_{p=0}^i\sum_{q=0}^j\hu^{p+q}\hv^{i+j-p-q+1}
C_i^pC_j^q(-1)^{j-q}(\xi_{p+1}\cdots\xi_{p+q}\zeta_{i-p+2}\cdots\zeta_{i+j-p-q+1})^{n+\frac{1}{2}-r}\\
&&
b_{i+1,j}(\xi_1,\ldots,\xi_p,\zeta_1,\ldots,\zeta_{i-p},\zeta_{i+1-p},\xi_{p+1},\ldots,\xi_{p+q},\zeta_{i-p+2},\ldots,
\zeta_{i+j-p-q+1}) \\
&=& \sum_{p=0}^{i+1}\sum_{q=0}^j\hu^{p+q}\hv^{i+1+j-p-q}
C_{i+1}^pC_j^q(-1)^{j-q}(\xi_{p+1}\cdots\xi_{p+q}\zeta_{i-p+2}\cdots\zeta_{i+1+j-p-q})^{n+\frac{1}{2}-r}\\
&&
b_{i+1,j}(\xi_1,\ldots,\xi_p,\zeta_1,\ldots,\zeta_{i+1-p},\xi_{p+1},\ldots,\xi_{p+q},\zeta_{i-p+2},\ldots,
\zeta_{i+1+j-p-q})
\end{eqnarray*}
and similarly we can prove the formula is valid for $j+1$. Finally, we need to symmetrise the
expression with respect to permutations of arguments $\xi$ and $\zeta$. \hfill $\diamond$

In the same manner, we can work out how a polynomial in $\hu$, $\hv$ and their derivatives changes
under the transformation $T^{-1}$. Here we only give the formula.
\begin{Pro}\label{Pro11}
Under the transformation $T^{-1}$, an arbitrary polynomial
${\hu}^i{\hv}^j {\hat
b}_{i,j}(\xi_1,\ldots,\xi_i,\zeta_1,\ldots,\zeta_j)$ takes the form
\begin{eqnarray}
\label{termuv} &&\hu^i \hv^j {\hat
b}_{i,j}(\xi_1,\ldots,\xi_i,\zeta_1,\ldots,\zeta_j)=\frac{1}{2^{i+j}}\sum_{p=0}^i\sum_{q=0}^j
u^{p+q}v^{i+j-p-q} C_i^pC_j^q(-1)^{j-q}\\ \nonumber &&
\langle\langle {\hat
b}_{i,j}(\xi_1,\ldots,\xi_p,\zeta_1,\ldots,\zeta_{i-p},\xi_{p+1},\ldots,\xi_{p+q},\zeta_{i-p+1},\ldots,\zeta_{i+j-p-q})
(\zeta_{1}\cdots\zeta_{i+j-p-q})^{-n-\frac{1}{2}+r}\rangle_{\xi}\rangle_{\zeta},
\end{eqnarray}
where $C_i^j$ are binomial coefficients.
\end{Pro}

These two propositions immediately lead to the following result.
\begin{Pro}\label{Pro2}
For any integer $s\geq 1$, if
$$\sum_{i=0}^s
u^iv^{s-i}b_{i,s-i}(\xi_1,\ldots,\xi_i,\zeta_1,\ldots,\zeta_{s-i})=
\sum_{l=0}^s\hu^l\hv^{s-l}\hat{b}_{l,s-l}(\xi_1,\ldots,\xi_l,\zeta_1,\ldots,\zeta_{s-l}),
$$
then
\begin{eqnarray}
\label{ha}
&&\hat{b}_{l,s-l}(\xi_1,\ldots,\xi_l,\zeta_1,\ldots,\zeta_{s-l})=\sum_{i=0}^s\sum_{p=max\{0,l-s+i\}}^{min\{i,l\}}
C_i^pC_{s-i}^{l-p}(-1)^{s-i-l+p}\\ \nonumber &&\langle\langle
b_{i,s-i}(\xi_1,\ldots,\xi_p,\zeta_1,\ldots,\zeta_{i-p},\xi_{p+1},\ldots,\xi_l,\zeta_{i-p+1},\ldots,\zeta_{s-l})
(\xi_{p+1}\cdots\xi_j\zeta_{i-p+1}\cdots\zeta_{s-l})^{n+\frac{1}{2}-r}\rangle_{\xi}\rangle_{\zeta}
\end{eqnarray}
and
\begin{eqnarray}\label{A}
&&b_{j,s-j}(\xi_1,\ldots,\xi_j,\zeta_1,\ldots,\zeta_{s-j})=\frac{1}{2^s}\sum_{i=0}^s\sum_{p=max\{0,j-s+i\}}^{min\{i,j\}}
C_i^pC_{s-i}^{j-p}(-1)^{s-i-j+p}\\ \nonumber &&\langle\langle
\hat{b}_{i,s-i}(\xi_1,\ldots,\xi_p,\zeta_1,\ldots,\zeta_{i-p},\xi_{p+1},\ldots,\xi_j,\zeta_{i-p+1},\ldots,\zeta_{s-j})
(\zeta_1\cdots\zeta_{s-j})^{-n-\frac{1}{2}+r}\rangle_{\xi}\rangle_{\zeta},
\end{eqnarray}
 where $C_i^j$ are binomial coefficients defined in Proposition
\ref{Pro1}.
\end{Pro}
{\bf Proof}. Applying formulae (\ref{termtr}) in Proposition
\ref{Pro1} to every term in the sum and collecting the coefficients
at every monomial $\hu^i\hv^j$, we obtain
\begin{eqnarray*}
&&\sum_{i=0}^su^iv^{s-i}b_{i,s-i}(\xi_1,\ldots,\xi_i,\zeta_1,\ldots,\zeta_{s-i})=\sum_{i=0}^s\sum_{p=0}^i\sum_{q=0}^{s-i}\hu^{p+q}\hv^{s-p-q}
C_i^pC_{s-i}^q(-1)^{s-i-q}\\  &&\langle\langle
b_{i,s-i}(\xi_1,\ldots,\xi_p,\zeta_1,\ldots,\zeta_{i-p},\xi_{p+1},\ldots,\xi_{p+q},\zeta_{i-p+1},\ldots,\zeta_{s-p-q})
(\xi_{p+1}\cdots\xi_{p+q}\zeta_{i-p+1}\cdots\zeta_{s-p-q})^{n+\frac{1}{2}-r}\rangle_{\xi}\rangle_{\zeta}
\\
&=&\sum_{i=0}^s\sum_{l=0}^s\sum_{p=max\{0,l-s+i\}}^{min\{i,\ l\}}\hu^{l}\hv^{s-l} C_i^pC_{s-i}^{l-p}(-1)^{s-i-l+p}\\
&&\langle\langle
b_{i,s-i}(\xi_1,\ldots,\xi_p,\zeta_1,\ldots,\zeta_{i-p},\xi_{p+1},\ldots,\xi_{l},\zeta_{i-p+1},\ldots,\zeta_{s-l})
(\xi_{p+1}\cdots\xi_{l}\zeta_{i-p+1}\cdots\zeta_{s-l})^{n+\frac{1}{2}-r}\rangle_{\xi}\rangle_{\zeta},
\end{eqnarray*}
which leads to formula (\ref{ha}). Similarly, using formulae (\ref
{termuv}) in Proposition \ref{Pro11}, we can prove formula
(\ref{A}). \hfill $\diamond$

Formula (\ref{ha}) in Proposition  \ref{Pro2} tells us how a
polynomial of degree $s$ changes under transformation $T$. The
following result is an immediate corollary.
\begin{Cor}\label{Cor1}
In variables $\hat{u}$ and $\hat{v}$, system (\ref{eqs}) takes the
form
\begin{eqnarray}
\left\{\begin{array}{l}\label{eqd}
\hu_t=\hu\xi_1^{n+\frac{1}{2}}+\frac{1}{2}\sum_{s\ge
2}\sum_{i=0}^s\hu^i\hv^{s-i}\frac{\hat{a}_{i,s-i}(\xi_1,\ldots,\xi_i,\zeta_1,\ldots,\zeta_{s-i})}{(\xi_1+\cdots+
\xi_i+\zeta_1+\cdots+\zeta_{s-i})^{n+\frac{1}{2}-r}}\\
\hv_t=-\hv\zeta_1^{n+\frac{1}{2}}-\frac{1}{2}\sum_{s\ge
2}\sum_{i=0}^s\hu^i\hv^{s-i}\frac{\hat{a}_{i,s-i}(\xi_1,\ldots,\xi_i,\zeta_1,\ldots,\zeta_{s-i})}{(\xi_1+\cdots+
\xi_i+\zeta_1+\cdots+\zeta_{s-i})^{n+\frac{1}{2}-r}}
\end{array}\right. ,
\end{eqnarray}
where
$\hat{a}_{i,s-i}(\xi_1,\ldots,\xi_i,\zeta_1,\ldots,\zeta_{s-i})$ are
defined in terms of
${a}_{j,s-j}(\xi_1,\ldots,\xi_j,\zeta_1,\ldots,\zeta_{s-j}),\,j=0,\ldots,s$
as in formula (\ref{ha}) in Proposition \ref{Pro2}.
\end{Cor}
{\bf Proof}. We perform transformation $T$  to system (\ref{eqs}):
the linear part becomes diagonal and we simply write the system as
\begin{eqnarray}
\left\{\begin{array}{l}\label{eqd0}
\hu_t=\hu\xi_1^{n+\frac{1}{2}}+\frac{1}{2\eta^{n+\frac{1}{2}-r}}\left(\sum_{s\ge
2}\sum_{i=0}^su^iv^{s-i}a_{i,s-i}(\xi_1,\ldots,\xi_i,\zeta_1,\ldots,\zeta_{s-i})\right)\\
\hv_t=-\hv\zeta_1^{n+\frac{1}{2}}-\frac{1}{2\eta^{n+\frac{1}{2}-r}}\left(\sum_{s\ge
2}\sum_{i=0}^su^iv^{s-i}a_{i,s-i}(\xi_1,\ldots,\xi_i,\zeta_1,\ldots,\zeta_{s-i})\right)
\end{array}\right. .
\end{eqnarray}
Applying Proposition \ref{Pro2}, we obtain system (\ref{eqd}) as
stated. \hfill $\diamond$

For the future references we explicitly write down the relation
between the quadratic terms of (\ref{eqs}) and of (\ref{eqd}):
\begin{eqnarray}
\label{a20h}
\hat{a}_{2,0}(\xi_1,\xi_2)&=&a_{2,0}(\xi_1,\xi_2)+a_{0,2}(\xi_1,\xi_2)(\xi_1\xi_2)^{n+\frac{1}{2}-r}+
\frac{1}{2}a_{1,1}(\xi_1,\xi_2)\xi_2^{n+\frac{1}{2}-r}+\frac{1}{2}a_{1,1}(\xi_2,\xi_1)
\xi_1^{n+\frac{1}{2}-r},\\ \label{a11h}
\hat{a}_{1,1}(\xi_1,\xi_2)&=&2a_{2,0}(\xi_1,\xi_2)-2a_{0,2}(\xi_1,\xi_2)(\xi_1\xi_2)^{n+\frac{1}{2}-r}
+a_{1,1}(\xi_2,\xi_1)\xi_1^{n+\frac{1}{2}-r}-
a_{1,1}(\xi_1,\xi_2)\xi_2^{n+\frac{1}{2}-r},\\ \label{a02h}
\hat{a}_{0,2}(\xi_1,\xi_2)&=&a_{2,0}(\xi_1,\xi_2)+a_{0,2}(\xi_1,\xi_2)(\xi_1\xi_2)^{n+\frac{1}{2}-r}-
\frac{1}{2}a_{1,1}(\xi_1,\xi_2)\xi_2^{n+\frac{1}{2}-r}-\frac{1}{2}a_{1,1}(\xi_2,\xi_1)\xi_1^{n+\frac{1}{2}-r}
\end{eqnarray}

We now express the symmetries (\ref{sym1}) and (\ref{sym2}) of
system (\ref{eqs}) in variables $\hu,\hv$.

Let us start with {\bf type I} symmetries (\ref{sym1}):
\[
u_{\tau}=u\xi_1^m+g,\qquad  \mbox{with} \quad g=\sum_{s\ge
2}\sum_{i=0}^s
u^iv^{s-i}A_{i,s-i}(\xi_1,\ldots,\xi_i,\zeta_1,\ldots,\zeta_{s-i}).
\]
Due to the compatibility condition (\ref{symd}), its second
component is of the form
\[
v_{\tau}=v\zeta_1^m+\frac{1}{\eta^r}\left(D_t(g)\right),
\]
where the action of the operator $D_t$ is defined in the symbolic
representation as
\[
D_t(g)=g_{*,u}(v\zeta_1^r)+g_{*,v}\left(u\xi_1^{2n+1-r}+\sum_{s\ge
2}\sum_{i=0}^su^iv^{s-i}a_{i,s-i}(\xi_1,\ldots,\xi_i,\zeta_1,\ldots,\zeta_{s-i})\right).
\]

In variables $\hu$ and $\hv$,  a {\bf type I} symmetry takes the
form
\begin{eqnarray}
\left\{\begin{array}{l} \label{sym1d}
\hu_{\tau}=\hu\xi_1^m+\frac{1}{2\eta^{n+\frac{1}{2}}}\left[\eta^{n+\frac{1}{2}}+\hat{D}_t\right](\hat{g})\\
\hv_{\tau}=\hv\zeta_1^m+\frac{1}{2\eta^{n+\frac{1}{2}}}\left[\eta^{n+\frac{1}{2}}-\hat{D}_t\right](\hat{g})
\end{array}\right. .
\end{eqnarray}
Here $\hat{g}$ stands for the transformed $g$ under transformation
$T$, whose exact expression can be obtained by applying Proposition
\ref{Pro2}. Notation $\hat{D}_t$ denotes the operator of total
derivation with respect to $t$ due to diagonal system (\ref{eqd}).
Hence
\[
\hat{D}_t(\hat{g})=\hat{g}_{*,\hu}(\hu_t)+\hat{g}_{*,\hv}(\hv_t).
\]

For {\bf type II} symmetries (\ref{sym2}):
\[
u_{\tau}=v\zeta_1^{m+r}+g,\qquad \mbox{with} \quad g=\sum_{s\ge
2}\sum_{i=0}^su^iv^{s-i}A_{i,s-i}(\xi_1,\ldots,\xi_i,\zeta_1,\ldots,\zeta_{s-i}),
\]
its second component is of the form
\[
v_{\tau}=u\xi_1^{2n+m+1-r}+\eta^m\left(\sum_{s\ge
2}\sum_{i=0}^su^iv^{s-i}a_{i,s-i}(\xi_1,\ldots,\xi_i,\zeta_1,\ldots,\zeta_{s-i})\right)+\eta^{-r}D_t(g).
\]

Using the same notations as described for {\bf type I} symmetries,
in variables $\hu$ and $\hv$ a {\bf type II} symmetry takes the form
\begin{eqnarray}
\left\{\begin{array}{l} \label{sym2d}
\hu_{\tau}=\hu\xi_1^{n+m+\frac{1}{2}}+\frac{1}{2\eta^{n+\frac{1}{2}}}\left[\eta^{n+\frac{1}{2}}+\hat{D}_t\right]\hat{g}+\frac{1}{2}
\eta^{m+r-n-\frac{1}{2}}\left(\sum_{s\ge
2}\sum_{i=0}^s\hu^i\hv^{s-i}\hat{a}_{i,s-i}(\xi_1,\cdots,\xi_i,\zeta_1,\cdots,\zeta_{s-i})\right)\\
\hv_{\tau}=-\hv\zeta_1^{n+m+\frac{1}{2}}+\frac{1}{2\eta^{n+\frac{1}{2}}}\left[\eta^{n+\frac{1}{2}}-\hat{D}_t\right]\hat{g}-
\frac{1}{2} \eta^{m+r-n-\frac{1}{2}}\left(\sum_{s\ge
2}\sum_{i=0}^s\hu^i\hv^{s-i}\hat{a}_{i,s-i}(\xi_1,\cdots,\xi_i,\zeta_1,\cdots,\zeta_{s-i})\right)
\end{array}.\right.
\end{eqnarray}

\subsection{Approximate symmetries and integrability conditions.}
We derive the compatibility conditions of (\ref{sym1}) or (\ref{sym2}) and (\ref{eqs}) via the
compatibility conditions of their transformed forms (\ref{sym1d}) or (\ref{sym2d}) and (\ref{eqd}). For
simplicity, we rewrite them in the following compact forms correspondingly:
\begin{eqnarray}
&&\left\{\begin{array}{l} \label{sys2}
u_{\tau}=u\Omega(\xi_1)+e^1\quad\quad\quad\quad e^j=\sum_{s\ge
2}\sum_{i=0}^su^iv^{s-i}e^j_{i,s-i}(\xi_1,\ldots,\xi_i,\zeta_1,\ldots,\zeta_{s-i}),\quad j=1,2\\
v_{\tau}=v\Omega(\zeta_1)+e^2
\end{array}\right.\\
&&\left\{\begin{array}{l} \label{sys3}
u_{\tau}=u\Omega(\xi_1)+e^1\\
v_{\tau}=-v\Omega(\zeta_1)+e^2
\end{array}\right.\\
&&\left\{\begin{array}{l} \label{sys1}
u_t=u\omega(\xi_1)+p\quad\quad\quad\quad p=\sum_{s\ge
2}\sum_{i=0}^su^iv^{s-i}p_{i,s-i}(\xi_1,\ldots,\xi_i,\zeta_1,\ldots,\zeta_{s-i}).\\
v_t=-v\omega(\zeta_1)-p
\end{array}\right.
\end{eqnarray}
Now we introduce a useful notation
\begin{eqnarray}\label{gfun}
{\cal G}_{i,s-i}^{\omega} (c_1,c_2;\xi_1,\cdots, \xi_i,\zeta_1,\cdots, \zeta_{s-i})=c_1
\omega(\sum_{l=1}^i \xi_l+\sum_{k=1}^{s-i} \zeta_k) -c_1 \sum_ {l=1}^i \omega(\xi_l)-c_2
\sum_{k=1}^{s-i} \omega(\zeta_k).
\end{eqnarray}

\begin{Pro}\label{Pro3}
System (\ref{sys2}) or system (\ref{sys3}) is compatible with
(\ref{sys1}) of degree $2$ if and only if the quadratic terms in
(\ref{sys2}) or system (\ref{sys3}) are related to the quadratic
terms in (\ref{sys1}) as follows:
\begin{eqnarray}
\nonumber &&e^1_{2,0}(\xi_1,\xi_2)=\frac{{\cal G}_{2,0}^{\Omega}
(1,c;\xi_1, \xi_2)} {{\cal G}_{2,0}^{\omega} (1,-1;\xi_1,
\xi_2)}p_{2,0}(\xi_1,\xi_2),\quad\quad
e^2_{2,0}(\xi_1,\xi_2)=-\frac{{\cal G}_{0,2}^{\Omega} (c,1;\xi_1,
\xi_2)} {{\cal G}_{0,2}^{\omega} (-1,1;\xi_1,
\xi_2)}p_{2,0}(\xi_1,\xi_2),\\ \nonumber
&&e^1_{1,1}(\xi_1,\zeta_1)=\frac{{\cal G}_{1,1}^{\Omega} (1,c;\xi_1,
\zeta_1)} {{\cal G}_{1,1}^{\omega} (1,-1;\xi_1,
\zeta_1)}p_{1,1}(\xi_1,\zeta_1),\quad\quad
e^2_{1,1}(\xi_1,\zeta_1)=-\frac{{\cal G}_{1,1}^{\Omega}
(c,1;\zeta_1, \xi_1)} {{\cal G}_{1,1}^{\omega} (-1,1;\zeta_1,
\xi_1)}p_{1,1}(\xi_1,\zeta_1),\\\nonumber
&&e^1_{0,2}(\zeta_1,\zeta_2)=\frac{{\cal G}_{0,2}^{\Omega}
(1,c;\zeta_1, \zeta_2)} {{\cal G}_{0,2}^{\omega} (1,-1;\zeta_1,
\zeta_2)}p_{0,2}(\zeta_1,\zeta_2),\quad\quad
e^2_{0,2}(\zeta_1,\zeta_2)=-\frac{{\cal G}_{2,0}^{\Omega}
(c,1;\zeta_1, \zeta_2)} {{\cal G}_{2,0}^{\omega} (-1,1;\zeta_1,
\zeta_2)}p_{0,2}(\zeta_1,\zeta_2),
\end{eqnarray}
where $c=1$ for the terms in (\ref{sys2}) and $c=-1$ for the terms
in (\ref{sys3}).
\end{Pro}

{\bf Proof}. Recall that for any polynomial pair
$$P=u^i v^{s-i}
a_{i,s-i}(\xi_1,\cdot, \xi_i,\zeta_1,\cdots, \zeta_{s-i})\quad
\mbox{and} \quad  Q=u^j v^{r-j} b_{j,r-j}(\xi_1,\cdot,
\xi_j,\zeta_1,\cdots, \zeta_{r-j}),$$ where $s\geq 2$, $r \geq 2$,
$0\leq i\leq s$ and $0\leq j\leq r$ are integers, we have  \cite{sw}
\begin{eqnarray}\label{G}
[\left(\begin{array}{c} c_1 u\omega (\xi_1)
\\ c_2 v\omega (\zeta_1) \end{array}\right)
,\left(\begin{array}{c}P\\Q \end{array}\right)] =\left(\begin{array}{c} {\cal G}_{i,s-i}^{\omega}
(c_1,c_2;\xi_1,\cdots, \xi_i,\zeta_1,\cdots, \zeta_{s-i})P \\
{\cal G}_{r-j,j}^{\omega} (c_2,c_1;\zeta_1,\cdots, \zeta_{r-j},\xi_1,\cdots, \xi_{j})Q
\end{array}\right)\, .
\end{eqnarray}

The compatibility conditions of (\ref{sys2}) and (\ref{sys1}) up to degree $2$ read as
\begin{eqnarray*}
&&[\left(\begin{array}{c} u\omega (\xi_1)
\\ - v\omega (\zeta_1) \end{array}\right)
,\left(\begin{array}{c}\sum_{i=0}^2 u^i v^{2-i}e_{i,2-i}^1(\xi_1,\cdots, \xi_i,\zeta_1, \cdots,
\zeta_{2-i})\\\sum_{i=0}^2 u^i v^{2-i}e_{i,2-i}^2(\xi_1,\cdots, \xi_i,\zeta_1, \cdots, \zeta_{2-i})
\end{array}\right)]\\
&=&[\left(\begin{array}{c} u\Omega (\xi_1)
\\ v\Omega (\zeta_1) \end{array}\right)
,\left(\begin{array}{c}\sum_{i=0}^2 u^i v^{2-i}p_{i,2-i}(\xi_1,\cdots, \xi_i,\zeta_1, \cdots,
\zeta_{2-i})\\ -\sum_{i=0}^2 u^i v^{2-i}p_{i,2-i}(\xi_1,\cdots, \xi_i,\zeta_1, \cdots,
\zeta_{2-i})\end{array}\right)]
\end{eqnarray*}
Using formula (\ref{G}) and collecting the terms at every monomial $u^iv^{2-i}$ for $i\in \{0,1,2\}$,
we obtain the formula for the quadratic terms in (\ref{sys2}) with $c=1$. Similarly, we can derive the
quadratic terms in (\ref{sys3}) as in the statement. \hfill $\diamond$

In fact, the results of Proposition \ref{Pro3} can be generated to
the terms of higher degree $i+j>2$, namely, one can derive that
\begin{eqnarray}
&&e^1_{i,j}(\xi_1,\ldots,\xi_{i},\zeta_1,\cdots,\zeta_{j})=\frac{{\cal
G}_{i,j}^{\Omega} (1,c;\xi_1,\cdots, \xi_i,\zeta_1,\cdots,
\zeta_{j}) } {{\cal G}_{i,j}^{\omega} (1,-1;\xi_1,\cdots,
\xi_i,\zeta_1,\cdots, \zeta_{j})
}p_{i,j}(\xi_1,\ldots,\xi_{i},\zeta_1,\cdots,\zeta_{j})+L_{i,j}^1\label{high1}\\
&&e^2_{i,j}(\xi_1,\ldots,\xi_{i},\zeta_1,\cdots,\zeta_{j})=-\frac{{\cal
G}_{j,i}^{\Omega} (c,1;\zeta_1,\cdots, \zeta_{j},\xi_1,\cdots,
\xi_{i})} {{\cal G}_{j,i}^{\omega} (-1,1;\zeta_1,\cdots,
\zeta_{j},\xi_1,\cdots,
\xi_{i})}p_{i,j}(\xi_1,\ldots,\xi_{i},\zeta_1,\cdots,\zeta_{j})+L_{i,j}^2
\label{high2},
\end{eqnarray}
where $c=1$ for the terms in (\ref{sys2}) and $c=-1$ for the terms in (\ref{sys3}). Expressions
$L_{i,j}^1$ and $L_{i,j}^2$ depend only on the terms of degree lower than $i+j$ in system (\ref{sys1}).
We will not write out them explicitly.

Now we are ready to derive the compatibility conditions of (\ref{sym1}) or (\ref{sym2}) and
(\ref{eqs}). Let us introduce the following notations:
\begin{eqnarray}
N_s^{(m)}(\xi_1,\ldots, \xi_s)&=&(\sum_{l=1}^s \xi_l)^m-\sum_{l=1}^s \xi_l^m\label{N}\\
S_{s,i}^{(n)}(\xi_1,\ldots,\xi_s)&=&(\xi_1+\cdots+\xi_s)^{2n+1}- \left(\sum_{l=1}^i
\xi_l^{n+\frac{1}{2}}- \sum_{l=i+1}^s \xi_{l}^{n+\frac{1}{2}}\right)^2,\label{S}\\
M_{s,i}^{(m)}(\xi_1,\ldots,\xi_s)&=&(\xi_1+\ldots+\xi_s)^m\left(\sum_{l=1}^i \xi_l^{n+\frac{1}{2}}-
\sum_{l=i+1}^s \xi_{l}^{n+\frac{1}{2}}\right)-\sum_{l=1}^i \xi_l^{n+m+\frac{1}{2}} +\sum_{l=i+1}^s
\xi_{l}^{n+m+\frac{1}{2}}.\label{M}
\end{eqnarray}

\begin{The}\label{th1}
Expression (\ref{sym1}) and expression (\ref{sym2}) are approximate symmetries of degree 2 of system
(\ref{eqs}) if and only if functions $A_{2,0},\,A_{1,1},\,A_{0,2}$ given by formulae
\begin{eqnarray}
\label{A20}
A_{2,0}(\xi_1,\xi_2)&=&\frac{1}{8}\left[2\hat{A}_{2,0}(\xi_1,\xi_2)+2\hat{A}_{0,2}(\xi_1,\xi_2)+
\hat{A}_{1,1}(\xi_1,\xi_2)+\hat{A}_{1,1}(\xi_2,\xi_1)\right],\\
\label{A11}
A_{1,1}(\xi_1,\xi_2)&=&\frac{1}{4\xi_2^{n+\frac{1}{2}-r}}\left[2\hat{A}_{2,0}(\xi_1,\xi_2)-2\hat{A}_{0,2}(\xi_1,\xi_2)-
\hat{A}_{1,1}(\xi_1,\xi_2)+\hat{A}_{1,1}(\xi_2,\xi_1) \right],\\
\label{A02}
A_{0,2}(\xi_1,\xi_2)&=&\frac{1}{8(\xi_1\xi_2)^{n+\frac{1}{2}-r}}
\left[2\hat{A}_{2,0}(\xi_1,\xi_2)+2\hat{A}_{0,2}(\xi_1,\xi_2)-
\hat{A}_{1,1}(\xi_1,\xi_2)-\hat{A}_{1,1}(\xi_2,\xi_1) \right].
\end{eqnarray}
are polynomials in $\xi_1,\xi_2$, where
$\hat{A}_{2,0},\hat{A}_{1,1},\hat{A}_{0,2}$ are determined as
follows:
\begin{enumerate}
\item For the quadratic terms in (\ref{sym1}),
\begin{eqnarray}
&&\hat{A}_{2,0}(\xi_1,\xi_2)=\frac{(\xi_1+\xi_2)^rN_2^{(m)}(\xi_1,\xi_2)}{S_{2,2}^{(n)}(\xi_1,\xi_2)}\hat{a}_{2,0}(\xi_1,\xi_2),\quad
\hat{A}_{1,1}(\xi_1,\xi_2)=\frac{(\xi_1+\xi_2)^rN_2^{(m)}(\xi_1,\xi_2)}{S_{2,1}^{(n)}(\xi_1,\xi_2)}\hat{a}_{1,1}(\xi_1,\xi_2),
\nonumber\\ &&
\hat{A}_{0,2}(\xi_1,\xi_2)=\frac{(\xi_1+\xi_2)^rN_2^{(m)}(\xi_1,\xi_2)}{S_{2,2}^{(n)}(\xi_1,\xi_2)}\hat{a}_{0,2}(\xi_1,\xi_2).
\label{Arel1}\end{eqnarray}
\item For the quadratic terms in (\ref{sym2}),
\begin{eqnarray}
&&\hat{A}_{2,0}(\xi_1,\xi_2)=\frac{(\xi_1+\xi_2)^rM_{2,2}^{(m)}(\xi_1,\xi_2)}{S_{2,2}^{(n)}(\xi_1,\xi_2)}\hat{a}_{2,0}(\xi_1,\xi_2),\
\hat{A}_{1,1}(\xi_1,\xi_2)=\frac{(\xi_1+\xi_2)^rM_{2,1}^{(m)}(\xi_1,\xi_2)}{S_{2,1}^{(n)}(\xi_1,\xi_2)}\hat{a}_{1,1}(\xi_1,\xi_2),
\nonumber\\
&&\hat{A}_{0,2}(\xi_1,\xi_2)=-\frac{(\xi_1+\xi_2)^rM_{2,2}^{(m)}(\xi_1,\xi_2)}{S_{2,2}^{(n)}(\xi_1,\xi_2)}\hat{a}_{0,2}(\xi_1,\xi_2).
\label{Arel2}\end{eqnarray}
\end{enumerate}
Here $\hat{a}_{2,0},\hat{a}_{1,1},\hat{a}_{0,2}$ are given in terms
of $a_{2,0},a_{1,1},a_{0,2}$ by (\ref{a20h}), (\ref{a11h}) and
(\ref{a02h}).
\end{The}

{\bf Proof}. Expression (\ref{sym1}) and (\ref{sym2}) are
approximate symmetries of degree $2$ of (\ref{eqs}) if and only if
they are compatible with (\ref{eqs}) up to degree $2$. Therefore the
diagonal forms of these symmetries (\ref{sym1d}) and (\ref{sym2d})
must be compatible with the diagonal form of our system (\ref{eqd}).
The latter is of the form of (\ref{sys1}) with
$\omega(x)=x^{n+\frac{1}{2}}$, while (\ref{sym1d}) and (\ref{sym2d})
are of the form (\ref{sys2}) with $\Omega(x)=x^m$ and (\ref{sys3})
with $\Omega(x)=x^{n+m+\frac{1}{2}}$ respectively. From Proposition
\ref{Pro2}, we know the relation of $A_{2,0},A_{1,1},A_{0,2}$ and
$\hat{A}_{2,0},\hat{A}_{1,1},\hat{A}_{0,2}$  as well as the relation
of $\hat{a}_{2,0},\hat{a}_{1,1},\hat{a}_{0,2}$ and
$a_{2,0},a_{1,1},a_{0,2}$. Now we need to show that
$\hat{A}_{2,0},\,\hat{A}_{1,1},\,\hat{A}_{0,2}$ are determined as
stated.

Consider first the case of {\bf type I} symmetries (\ref{sym1}).
Applying Proposition \ref{Pro3} in the case of system (\ref{sys1})
and (\ref{sys2}) with $\omega(x)=x^{n+\frac{1}{2}}$ and
$\Omega(x)=x^m$, we obtain (for the coefficient at $u^2v^0$):
$$
e^1_{2,0}(\xi_1,\xi_2)=\frac{N_2^{(m)}(\xi_1, \xi_2)}
{(\xi_1+\xi_2)^{n+\frac{1}{2}}-\xi_1^{n+\frac{1}{2}}-\xi_2^{n+\frac{1}{2}}}p_{2,0}(\xi_1,\xi_2).
$$
Comparing system (\ref{sys2}) to system (\ref{sym1d}) and system
(\ref{sys1}) to system (\ref{eqd}), we have
\begin{eqnarray*}
e^1_{2,0}(\xi_1,\xi_2)=\frac{(\xi_1+\xi_2)^{n+\frac{1}{2}}+\xi_1^{n+\frac{1}{2}}+\xi_2^{n+\frac{1}{2}}}
{2(\xi_1+\xi_2)^{n+\frac{1}{2}}} \hat{A}_{2,0}(\xi_1,\xi_2)\quad
\mbox{and} \quad
p_{2,0}(\xi_1,\xi_2)=\frac{\hat{a}_{2,0}(\xi_1,\xi_2)}{2(\xi_1+\xi_2)^{n+\frac{1}{2}-r}}.
\end{eqnarray*}
Substituting these into the previous formula, we get
\[
\hat{A}_{2,0}(\xi_1,\xi_2)=\frac{(\xi_1+\xi_2)^rN_2^{(m)}(\xi_1,\xi_2)}
{S_{2,2}^{(n)}(\xi_1,\xi_2)}\hat{a}_{2,0}(\xi_1,\xi_2),
\]
the first formula of (\ref{Arel1}). The other two relations can be
obtained in the same way.

Consider now the case of {\bf type II} symmetries (\ref{sym2}).
Using Proposition \ref{Pro3} in the case of systems (\ref{sys1}) and
(\ref{sys3}) with $\omega(x)=x^{n+\frac{1}{2}}$ and
$\Omega(x)=x^{n+m+\frac{1}{2}}$ we find (for the coefficient at
$u^2v^0$):
$$
e^1_{2,0}(\xi_1,\xi_2)=\frac{(\xi_1+\xi_2)^{m+n+\frac{1}{2}}-\xi_1^{m+n+\frac{1}{2}}-\xi_2^{m+n+\frac{1}{2}}}
{(\xi_1+\xi_2)^{n+\frac{1}{2}}-\xi_1^{n+\frac{1}{2}}-\xi_2^{n+\frac{1}{2}}}p_{2,0}(\xi_1,\xi_2).
$$
Comparing system (\ref{sys3}) to system (\ref{sym2d}) , we know
\begin{eqnarray*}
e^1_{2,0}(\xi_1,\xi_2)=\frac{(\xi_1+\xi_2)^{n+\frac{1}{2}}+\xi_1^{n+\frac{1}{2}}+\xi_2^{n+\frac{1}{2}}}{2(\xi_1+\xi_2)^{n+\frac{1}{2}}}
\hat{A}_{2,0}(\xi_1,\xi_2)+\frac{(\xi_1+\xi_2)^{m+r}}{2(\xi_1+\xi_2)^{n+\frac{1}{2}}}\hat{a}_{2,0}(\xi_1,\xi_2).
\end{eqnarray*}
This leads to
\[
\hat{A}_{2,0}(\xi_1,\xi_2)=\frac{(\xi_1+\xi_2)^rM_{2,2}^{(m)}(\xi_1,\xi_2)}
{S_{2,2}^{(n)}(\xi_1,\xi_2)}\hat{a}_{2,0}(\xi_1,\xi_2),
\]
the first formula of (\ref{Arel2}). The other two formula can be
obtained similarly. \hfill $\diamond$

The study of symmetries of (\ref{eqs}) is the study of the
divisibility of certain special functions. If we introduce variables
$x_1,x_2$ instead of $\xi_1,\xi_2$ by relations
\[
x_1=\xi_1^{\frac{1}{2}},\quad x_2=\xi_2^{\frac{1}{2}}
\]
then functions $\hat{a}_{2,0}(x_1^2,x_2^2),\ \hat{a}_{1,1}(x_1^2,x_2^2),\ \hat{a}_{0,2}(x_1^2,x_2^2),\
S_{2,i}^{(n)}(x_1^2,x_2^2)$ and $M_{2,i}^{(m)}(x_1^2,x_2^2)$ become {\it polynomials } in $x_1,x_2$
(see formulae (\ref{a20h}), (\ref{a11h}), (\ref{a02h}), (\ref{S}) and (\ref{M}) ).

\begin{Cor}\label{Cor2}
If (\ref{sym1}) and (\ref{sym2}) are approximate symmetries of
(\ref{eqs}) of degree 2 then functions $\hat{A}_{2,0}(x_1^2,x_2^2),\
\hat{A}_{1,1}(x_1^2,x_2^2),$ and $ \hat{A}_{0,2}(x_1^2,x_2^2)$
defined by (\ref{Arel1}) and (\ref{Arel2}) are polynomials in
$x_1,x_2$.
\end{Cor}

{\bf Proof}. From Theorem \ref{th1}, if (\ref{sym1}) and (\ref{sym2}) are approximate symmetries of
(\ref{eqs}) of degree 2 then $A_{2,0}(x_1^2,x_2^2),\ A_{1,1}(x_1^2,x_2^2),$ and $A_{0,2}(x_1^2,x_2^2)$
are polynomials in $x_1,x_2$ (in fact these functions are polynomials in $x_1^2,x_2^2$). We know
$\hat{A}_{2,0}(x_1^2,x_2^2), \hat{A}_{1,1}(x_1^2,x_2^2), \hat{A}_{0,2}(x_1^2,x_2^2)$ can be expressed
as polynomials in $A_{2,0},A_{1,1},A_{0,2}$ with coefficients being polynomials in $x_1,x_2$ according
to Proposition \ref{Pro2}. Therefore $\hat{A}_{2,0},\ \hat{A}_{1,1},$ and $\hat{A}_{0,2}$ are
polynomials in $x_1,x_2$. \hfill $\diamond$

We are able to obtain explicit recursive relations for determining
the higher degree terms in (\ref{sym1}) and (\ref{sym2}). Similar as
formula (\ref{high1}) and (\ref{high2}), we present the strictures
of these expressions without the full details. From the proof of
Theorem \ref{th1}, we only need to show how functions
$\hat{A}_{i,s-i}$ are related to functions $\hat{a}_{i,r-i}$, where
$r\leq s$ and these can be obtained by using formula (\ref{high1})
and (\ref{high2}) and comparing the notations in (\ref{sys2}),
(\ref{sys3}) and (\ref{sys1}) with the ones in (\ref{sym1d}),
(\ref{sym2d}) and (\ref{eqd}).

Suppose that  (\ref{sym1}) or (\ref{sym2}) is an approximate
symmetry of (\ref{eqs}) of degree $p\ge 3$. Then for any $3\le s\le
p$ and any $i=0,\ldots, s$ we have
\begin{eqnarray}
&&\mbox{{\bf Type I}}: \quad \hat{A}_{i,s-i}(\xi_1,\ldots,\xi_s)=\frac{(\sum_{k=1}^s \xi_k)^r
N_s^{(m)}(\xi_1,\ldots,\xi_s)}
{S_{s,i}^{(n)}(\xi_1,\ldots,\xi_s)}\hat{a}_{i,s-i}(\xi_1,\ldots,\xi_s)+L_i^{(s)},\label{AhigherI} \\
&&\mbox{{\bf Type II}}:\quad
\hat{A}_{i,s-i}(\xi_1,\ldots,\xi_s)=\frac{(\sum_{k=1}^s\xi_k)^rM_{s,i}^{(m)}(\xi_1,\ldots,\xi_s)}
{S_{s,i}^{(n)}(\xi_1,\ldots,\xi_s)}\hat{a}_{i,s-i}(\xi_1,\ldots,\xi_s)+\tilde{L}_i^{(s)},\label{AhigherII}
\end{eqnarray}
where functions $\hat{A}_{i,s-i}$ and $\hat{a}_{i,s-i}$ are given in terms of $A_{i,s-i}$ and
$a_{i,s-i}$ according to Proposition \ref{Pro2}; $L^{(s)}_i$ and $\tilde{L}_i^{(s)}$ depend only on
lower degree terms $a_{i,j}, \,i+j<s$, and $N_s^{(m)}$, $S_{s,i}^{(n)}$ and $M_{s,i}^{(m)}$ are defined
as (\ref{N}), (\ref{S}) and (\ref{M}).

Again if we introduce variables
$$x_1=\xi_1^{\frac{1}{2}},\ldots,x_s=\xi_s^{\frac{1}{2}},$$
functions $\hat{a}_{i,s-i}(x_1^2,\ldots,x_s^2),\ \hat{A}_{i,s-i}(x_1^2,\ldots,x_s^2),\
S_{s,i}^{(n)}(x_1^2,\ldots,x_s^2)$, and $M_{s,i}^{(m)}(x_1^2,\ldots,x_s^2)$ all become {\it
polynomials} in variables $x_1,\ldots,x_s$.

\section{Global classification result.}\label{Sec4}
In this section, we state and prove the global classification result
of system (\ref{eq2}) in the sense that we list out all the
integrable equations of arbitrary order. We begin with an crucial
theorem leading to global classification. In order to prove the
theorem, we prove the irreducibility of a family of special
polynomials.

\begin{Pro}\label{Pro4}
For any $i=0,\ldots,s$, polynomials
$$
S_{s,i}^{(n)}(x_1^2,\ldots,x_s^2)=(x_1^2+\cdots+x_s^2)^{2n+1}-\left(x_1^{2n+1}+\cdots+x_i^{2n+1}-
x_{i+1}^{2n+1}-\cdots-x_s^{2n+1} \right)^2
$$
are irreducible polynomials in $x_1,\ldots,x_s$ over $\bbbc$ when
$s\ge 3$ and $n\ge 1$.
\end{Pro}

{\bf Proof}. It suffices to prove that the polynomial $S_{s,s}^{(n)}(x_1^2,\ldots,x_s^2)$, cf.
(\ref{S}), is irreducible over $\bbbc$ since all the rest  can be obtained from it by taking negation
of $s-i$ arguments $x_{i+1},\ldots x_{s}$.

Let us first suppose that $s\geq 4$. If $S_{s,s}^{(n)}$ is reducible then the projective hyperspace
given by $S_{s,s}^{(n)}=0$ consists of two components. These components intersect in infinite number of
points, which should be singularities of the hyperspace. These singular points can be computed by
setting the derivatives of $S_{s,s}^{(n)}$ with respect to $x_j,\,j=1,\ldots,s$ equal to zero.  We
obtain:
\[
\frac{\partial S_{s,s}^{(n)}}{\partial
x_j}=2(2n+1)x_j\left[x_j^{2n-1}(x_1^{2n+1}+\cdots+x_s^{2n+1})-(x_1^2+\cdots+x_s^2)^{2n}\right]=0,\,\,j=1,\ldots,s.
\]
So the singular points are either the solutions
\begin{eqnarray}\label{t1}
x_1^2+\cdots+x_s^2=0,\quad x_1^{2n+1}+\cdots+x_s^{2n+1}=0
\end{eqnarray}
or the solutions of
\begin{eqnarray}\label{t2}
x_1^{2n-1}=x_2^{2n-1}=\cdots=x_s^{2n-1}=x,\quad
(x_1^2+\cdots+x_s^2)^{2n-1}=x^2.
\end{eqnarray}
The singular points from (\ref{t1}) are kink points, which do not
contribute to factorization. From (\ref{t2}), it follows that
\begin{eqnarray}\label{t3}
(x_1^2)^{2n-1}=(x_2^2)^{2n-1}=\cdots=(x_s^2)^{2n-1}=(x_1^2+\cdots+x_s^2)^{2n-1}.
\end{eqnarray}
Hence the coordinates differ by $(4n-2)$-th root of unity. Thus we get finitely many singular points.
Thus the assumption that $S_{s,s}^{(n)}$ is reducible is false.

When $s=3$, system (\ref{t3}) writes as
\[
(x_1^2)^{2n-1}=(x_2^2)^{2n-1}=(x_3^2)^{2n-1}=(x_1^2+x_2^2+x_3^2)^{2n-1}.
\]
By taking $x_3=1$, we have that $x_1^2$, $x_2^2$ and $w$ are $2n-1$-st roots of unity such that
$w=x_1^2+x_2^2+1$. Note that four complex numbers of the same absolute value can only add up to zero if
they form the sides of a parallelogram with equal sides. Therefore either $w=1$ or $x_i^2=-1$, $i=1,2$.
If $w=1$, then $x_1^2=-x_2^2$ leading to $(x_1^2)^{2n-1}=-(x_2^2)^{2n-1}$, contradicting to
$(x_1^2)^{2n-1}=(x_2^2)^{2n-1}$. If $x_i^2=-1$, we have $(x_i^2)^{2n-1}=-1$, contradicting to
$(x_2^2)^{2n-1}=(x_3^2)^{2n-1}=1$. Hence, $S_{3,3}^{(n)}$ is irreducible. \hfill $\diamond$

The irreducibility of this family of polynomials leads to the
following useful result in testing integrability.
\begin{Cor}\label{Cor3}
If there are no quadratic terms in homogeneous system (\ref{eqs})
with $W(u)>0$, i.e.
\[
a_{2,0}(\xi_1,\xi_2)=a_{1,1}(\xi_1,\zeta_1)=
a_{0,2}(\zeta_1,\zeta_2)=0,
\]
and $a_{i,j}\ne 0$ for some $i,j,\,i+j>2$, then system (\ref{eqs})
does not possess higher symmetries of any order.
\end{Cor}

{\bf Proof}. Without loss of generality, we assume that some of the
terms in (\ref{eqs}) of degree $s\geq 3$
are not equal zero and all terms of degree less then $s$ are equal zero. Suppose that such system
possesses a type I symmetry (\ref{sym1}) of order $m>1$. Then it is easy to see from formulae
(\ref{Arel1}) and (\ref{AhigherI}) that $A_{i,j}(\xi_1,\ldots,\xi_i,\zeta_1,\ldots,\zeta_j)=0,\,i+j<s$
and
\[
\hat{A}_{i,s-i}(x_1^2,\ldots,x_s^2)=\frac{(x_1^2+\cdots+x_s^2)^r N_s^{(m)}(x_1^2,\ldots,x_s^2)}
{S_{s,i}^{(n)}(x_1^2,\ldots,x_s^2)}\hat{a}_{i,s-i}(x_1^2,\ldots,x_s^2),\,\,i=0,\ldots,s.
\]
The left hand side of the above formula must be polynomial in $x_1,\ldots,x_s$. From Proposition
\ref{Pro4} we know that polynomials $S_{s,i}^{(n)}(x_1^2,\ldots,x_s^2),\,\,i=0,\ldots,s$ are
irreducible polynomials over $\bbbc$ and these polynomials do not divide
$$(x_1^2+\cdots+x_s^2)^rN_s^{(m)}(x_1^2,\ldots,x_s^2)$$
 for any $m>1$ and $r\ge 0$. Therefore
$S_{s,i}^{(n)}$ must divide $\hat{a}_{i,s-i}(x_1^2,\ldots,x_s^2)$, what is impossible since
\[
\deg(\hat{a}_{i,s-i}(x_1^2,\ldots,x_s^2))=4n+2-2(s-1)W(u)<\deg(S_{s,i}^{(n)}(x_1^2,\ldots,x_s^2))=4n+2,\quad
i=0,\ldots,s.
\]
Thus system (\ref{eqs}) does not possess any type I symmetry. The
consideration in the case of type II symmetries is similar.
\hfill$\diamond$

We are now ready to prove the following important theorem in
classification of integrable homogeneous systems of the form
(\ref{eqs}).
\begin{The}\label{Th1}
Consider homogeneous system (\ref{eqs}) with $W(u)>0$. Assume that
it possesses an $m$-th order { type I} or { type II} symmetry of the
form (\ref{sym1}) or (\ref{sym2}). Suppose there is another system
of the same weight and of the same form
\begin{eqnarray}
\left\{\begin{array}{l}\label{eqb} u_t=v\zeta_1^r,\\
v_t=u\xi_1^{2n+1-r}+\sum_{k\ge 2}\sum_{i=0}^k
u^iv^{k-i}b_{i,k-i}(\xi_1,\ldots,\xi_k,\zeta_1,\ldots,\zeta_{k-i}),
\end{array}\right.
\end{eqnarray}
whose quadratic terms equal to those of (\ref{eqs}), that is,
$b_{i,2-i}(x,y)=a_{i,2-i}(x,y),\,i=0,1,2$. Then if system
(\ref{eqb}) possesses an $m$-th order type I or type II symmetry,
then equation (\ref{eqb}) and (\ref{eqs}) are equal and sharing the
same type I or type II symmetry of order $m$.
\end{The}

{\bf Proof}. We will prove the statement by induction. Consider
 the case of symmetries of type I. Let us suppose that
equation (\ref{eqb}) possesses a {type I} symmetry of the form
\begin{equation}
\label{symb} u_{\tau}=u\xi_1^m+\sum_{s\ge 2}\sum_{j=0}^s
u^jv^{s-j}B_{j,s-j}(\xi_1,\ldots,\xi_j,\zeta_1,\ldots,\zeta_{s-j})
\end{equation}
The quadratic  terms of this symmetry coincide with those of
(\ref{sym1}) $B_{i,2-i}(x,y)=A_{i,2-i}(x,y),\,i=0,1,2$ because the
quadratic terms of (\ref{eqb}) and (\ref{eqs}) coincide. Let us
prove that the terms of degree $3$ in (\ref{eqs}) and (\ref{eqb})
coincide. According to formula (\ref{AhigherI}) we have
\begin{eqnarray}
\nonumber \hat{A}_{i,3-i}(x_1^2,x_2^2,x_3^2)=\frac{(x_1^2+x_2^2+x_3^2)^r N_3^{(m)}(x_1^2,x_2^2,x_3^2)}
{S_{3,i}^{(n)}(x_1^2,x_2^2,x_3^2)}\hat{a}_{i,3-i}(x_1^2,x_2^2,x_3^2)+L_i^{(3)},\,\, i=0,1,2,3
\end{eqnarray}
for the coefficients of the symmetry (\ref{sym1}) of (\ref{eqs}) and
\begin{eqnarray}
\nonumber \hat{B}_{i,3-i}(x_1^2,x_2^2,x_3^2)=\frac{(x_1^2+x_2^2+x_3^2)^r N_3^{(m)}(x_1^2,x_2^2,x_3^2)}
{S_{3,i}^{(n)}(x_1^2,x_2^2,x_3^2)}\hat{b}_{i,3-i}(x_1^2,x_2^2,x_3^2)+\tilde{L}_i^{(3)},\,\, i=0,1,2,3
\end{eqnarray}
for the coefficients of the symmetry (\ref{symb}) of (\ref{eqb}).

Notice that ${L}^{(3)}_i={\tilde{L}}^{(3)}_i,\,i=0,1,2,3$ since they
only depend on the linear and quadratic terms of equations
(\ref{eqs}), (\ref{eqb}) and their symmetries (\ref{sym1}) and
(\ref{symb}), which are equal by the assumption. Therefore
\begin{eqnarray}
\nonumber
&&\hat{B}_{i,3-i}(x_1^2,x_2^2,x_3^2)-\hat{A}_{i,3-i}(x_1^2,x_2^2,x_3^2)=\\
\nonumber\\ \nonumber &&=\frac{(x_1^2+x_2^2+x_3^2)^r N_3^{(m)}(x_1^2,x_2^2,x_3^2)}
{S_{3,i}^{(n)}(x_1^2,x_2^2,x_3^2)}
\left(\hat{b}_{i,3-i}(x_1^2,x_2^2,x_3^2)-\hat{a}_{i,3-i}(x_1^2,x_2^2,x_3^2)\right).
\end{eqnarray}
Its left hand side must be a polynomial in $x_1,x_2,x_3$. In proposition \ref{Pro4} we proved that
polynomials $S_{3,i}^{(n)}(x_1^2,x_2^2,x_3^2)$ are irreducible polynomials over $\bbbc$. It is easy to
see that it does not divide $(x_1^2+x_2^2+x_3^2)^r N_3^{(m)}(x_1^2,x_2^2,x_3^2)$ for any $n\ge 1$,
$m>1$ and $i=0,1,2,3$. Therefore $S_{3,i}^{(n)}(x_1^2,x_2^2,x_3^2)$ must divide
$\hat{b}_{i,3-i}(x_1^2,x_2^2,x_3^2)-\hat{a}_{i,3-i}(x_1^2,x_2^2,x_3^2)$. But this is impossible since,
for $i\in \{0,1,2,3\}$,
\[
\deg(\hat{b}_{i,3-i}(x_1^2,x_2^2,x_3^2)-\hat{a}_{i,3-i}(x_1^2,x_2^2,x_3^2))=2(2n+1-2W(u)-r)<
\deg(S_{3,i}^{(n)}(x_1^2,x_2^2,x_3^2))=4n+2
\]
due to the assumption $W(u)>0$ and $r\ge 0$. Hence
$$\hat{b}_{i,3-i}(x_1^2,x_2^2,x_3^2)-\hat{a}_{i,3-i}(x_1^2,x_2^2,x_3^2)=0,\quad
\hat{B}_{i,3-i}(x_1^2,x_2^2,x_3^2)-\hat{A}_{i,3-i}(x_1^2,x_2^2,x_3^2)=0,
$$ that is,
$b_{i,3-i}(\xi_1,\xi_2,\xi_3)=a_{i,3-i}(\xi_1,\xi_2,\xi_3),\,\,B_{i,3-i}(\xi_1,\xi_2,\xi_3)=A_{i,3-i}(\xi_1,\xi_2,\xi_3)
$.

Suppose we have proved that the terms of degree $s-1$ coincide. For
terms of degree $s$, we have, according to formula (\ref{AhigherI}),
\begin{eqnarray}
\nonumber \hat{A}_{i,s-i}(x_1^2,\ldots,x_s^2)=\frac{(x_1^2+\cdots+x_s^2)^r
N_s^{(m)}(\xi_1,\ldots,\xi_s)}
{S_{s,i}^{(n)}(\xi_1,\ldots,\xi_s)}\hat{a}_{i,s-i}(x_1^2,\ldots,x_s^2)+L_i^{(s)},\,\, i=0,\ldots,s
\end{eqnarray}
for the coefficients of the symmetry (\ref{sym1}) of (\ref{eqs}) and
\begin{eqnarray}
\nonumber \hat{B}_{i,s-i}(x_1^2,\ldots,x_s^2)=\frac{(x_1^2+\cdots+x_s^2)^r
N_s^{(m)}(\xi_1,\ldots,\xi_s)}
{S_{s,i}^{(n)}(\xi_1,\ldots,\xi_s)}\hat{b}_{i,s-i}(x_1^2,\ldots,x_s^2)+\tilde{L}_i^{(s)},\,\,
i=0,\ldots,s
\end{eqnarray}
for the coefficients of the symmetry (\ref{symb}) of (\ref{eqb}).
Terms $L_i^{(s)}$ and $\tilde{L}_i^{(s)}$ depend only on
coefficients of degree less than $s$ and therefore they are equal
$L_i^{(s)}=\tilde{L}_i^{(s)},\,i=0,\ldots,s$. Hence
\begin{eqnarray*}
&&\hat{B}_{i,s-i}(x_1^2,\ldots,x_s^2)-\hat{A}_{i,s-i}(x_1^2,\ldots,x_s^2)
=\frac{(x_1^2+\cdots+x_s^2)^rN_s^{(m)}(\xi_1,\ldots,\xi_s)}
{S_{s,i}^{(n)}(\xi_1,\ldots,\xi_s)}\left(\hat{b}_{i,s-i}(x_1^2,\ldots,x_s^2)-\hat{a}_{i,s-i}(x_1^2,\ldots,x_s^2)\right).
\end{eqnarray*}
Again the left hand side of the last formula must be a polynomial in $x_1,\ldots,x_s$ and therefore
$S_{s,i}^{(n)}(x_1^2,\ldots,x_s^2)$ must divide
$\hat{b}_{i,s-i}(x_1^2,\ldots,x_s^2)-\hat{a}_{i,s-i}(x_1^2,\ldots,x_s^2)$. This is impossible since
\[
\deg(\hat{b}_{i,s-i}(x_1^2,\ldots,x_s^2)-\hat{a}_{i,s-i}(x_1^2,\ldots,x_s^2))=4n+2-2(s-1)W(u)<
\deg(S_{s,i}^{(n)}(x_1^2,\ldots,x_s^2))=4n+2,\,i=0,\ldots,s
\]
using the assumption $W(u)>0$. Hence $b_{i,s-i}=a_{i,s-i}$ and
$B_{i,s-i}=A_{i,s-i}$ for $i=0,\ldots,s$. Thus we proved the
statement for type I symmetries. The proof in the case of symmetries
of type II is analogous. \hfill$\diamond$

Finally, we state and prove the global classification result of
homogeneous system (\ref{eq2}).

\begin{The}\label{Th2}
If a homogeneous system (\ref{eq2}) with $W(u)=w>0$ possesses a hierarchy  of infinitely many higher
symmetries, then it is one of the systems in the following list up to re-scaling $u\to\alpha
u,v\to\beta v,t\to\gamma t,x\to\delta x,$ where $\alpha,\beta,\gamma,\delta$ are constant:
\begin{eqnarray}
&&\left\{\begin{array}{l}\label{eqS2} u_t=v_1,\\
v_t=u_2+3uv_1+vu_1-3u^2u_1,
\end{array}\right.\\
&&\left\{\begin{array}{l}\label{Burg} u_t=v_1,\\
v_t=(D_x+u)^{2n}(u)-v^2,\quad n=1,2,3,\ldots .
\end{array}\right.
\end{eqnarray}
\end{The}

{\bf Proof}. We prove the theorem in symbolic representation. We know that system (\ref{eq2}) in
symbolic representation takes the form of (\ref{eqs}). In Corollary \ref{Cor3} we have proved that if
(\ref{eq2}) possesses higher symmetries then the differential polynomial $F$ in (\ref{eq2}) must
contain quadratic terms and therefore $w=W(u)>0$ is either integer or half-integer. Indeed,
\begin{eqnarray*}
\ring^2_{2n+1-r+w}&=&\mbox{span}\{u_iu_j|i+j=2n+1-r-w,\,i,j\in\bbbn_0\}\bigoplus\\
&&\mbox{span}\{u_iv_j|i+j=n+\frac{1}{2}-w,\,i,j\in\bbbn_0\}\bigoplus\mbox{span}\{v_iv_j|i+j=r-w,\,i,j\in\bbbn_0\},
\end{eqnarray*}
where $\bbbn_0=\bbbn \bigcup \{0\}$. Therefore if differential polynomial $F$ in (\ref{eq2}) contains
quadratic terms then the weight of variable $u$ is either integer or half-integer, i.e. either
$w\in\bbbn$ or $w=s-\frac{1}{2},\,s\in\bbbn$. Furthermore, from the Theorem \ref{Th1} it follows that
it suffice to classify integrable systems (\ref{eq2}) up to quadratic terms. If system (\ref{eq2}) or
rather say (\ref{eqs}) possesses a hierarchy of infinitely many higher symmetries, it possesses
infinitely many approximate symmetries of degree $2$ of type I or type II. We first assume that these
are of type I of the form
\[
u_{\tau_j}=u\xi_1^{m_j}+u^2A_{2,0}^{(j)}(\xi_1,\xi_2)+uv
A_{1,1}^{(j)}(\xi_1,\zeta_1)+v^2A_{0,2}^{(j)}(\zeta_1,\zeta_2)+o(\ring^2),\quad
j=1,2,3,\ldots
\]
where $1<m_1<m_2<\cdots<m_j<m_{j+1}<\cdots $. From Theorem \ref{th1}
and Corollary \ref{Cor2}, it follows that
\begin{eqnarray*}
&&\hat{A}^{(j)}_{2,0}(x_1^2,x_2^2)=(x_1^2+x_2^2)^r\frac{{N}_2^{(m_j)}(x_1^2,x_2^2)}{{S}_{2,2}^{(n)}(x_1^2,x_2^2)}
\hat{a}_{2,0}(x_1^2,x_2^2),\quad
\hat{A}^{(j)}_{1,1}(x_1^2,x_2^2)=(x_1^2+x_2^2)^r\frac{N_2^{(m_j)}(x_1^2,x_2^2)}{{S}_{2,1}^{(n)}(x_1^2,x_2^2)}
\hat{a}_{1,1}(x_1^2,x_2^2)\\
&&\hat{A}^{(j)}_{0,2}(x_1^2,x_2^2)=(x_1^2+x_2^2)^r\frac{{N}_2^{(m_j)}(x_1^2,x_2^2)}{{S}_{2,2}^{(n)}(x_1^2,x_2^2)}
\hat{a}_{0,2}(x_1^2,x_2^2)
\end{eqnarray*}
must be polynomials in $x_1,x_2$ for any $m_j,\ j=1,2,\ldots$. Notice that polynomials
$S_{2,i}^{(n)}(x_1^2,x_2^2)$ and $N_2^{(m_j)}(x_1^2,x_2^2)$ can be factorized as
\[
S_{2,i}^{(n)}(x_1^2,x_2^2)=x_1^2x_2^2\hat{S}_{2,i}^{(n)}(x_1^2,x_2^2),\quad
N_2^{(m_j)}(x_1^2,x_2^2)=x_1^2x_2^2\hat{N}_2^{(m_j)}(x_1^2,x_2^2),
\]
which leads to
\begin{eqnarray*}
&&\hat{A}^{(j)}_{2,0}(x_1^2,x_2^2)=(x_1^2+x_2^2)^r\frac{\hat{N}_2^{(m_j)}(x_1^2,x_2^2)}{\hat{S}_{2,2}^{(n)}(x_1^2,x_2^2)}
\hat{a}_{2,0}(x_1^2,x_2^2),\quad
\hat{A}^{(j)}_{1,1}(x_1^2,x_2^2)=(x_1^2+x_2^2)^r\frac{\hat{N}_2^{(m_j)}(x_1^2,x_2^2)}{\hat{S}_{2,1}^{(n)}(x_1^2,x_2^2)}
\hat{a}_{1,1}(x_1^2,x_2^2)\\
&&\hat{A}^{(j)}_{0,2}(x_1^2,x_2^2)=(x_1^2+x_2^2)^r\frac{\hat{N}_2^{(m_j)}(x_1^2,x_2^2)}{\hat{S}_{2,2}^{(n)}(x_1^2,x_2^2)}
\hat{a}_{0,2}(x_1^2,x_2^2).
\end{eqnarray*}

We can easily check $\gcd((x_1^2+x_2^2)^r,S_{2,i}^{(n)}(x_1^2,x_2^2))
=\gcd((x_1^2+x_2^2)^r,\hat{S}_{2,i}^{(n)}(x_1^2,x_2^2))=1$. Moreover, from Lemma \ref{Lem1} in Appendix
A it follows that if there exist infinitely many $m_j$ such that $\hat{A}_{i,2-i},\,\,i=0,1,2$  are
polynomial, then $m_j=2,3,4,\ldots $ and
\begin{eqnarray}\label{div}
\hat{S}_{2,2}^{(n)}(x_1^2,x_2^2)|\hat{a}_{2,0}(x_1^2,x_2^2),\quad
\hat{S}_{2,1}^{(n)}(x_1^2,x_2^2)|\hat{a}_{1,1}(x_1^2,x_2^2),\quad
\hat{S}_{2,2}^{(n)}(x_1^2,x_2^2)|\hat{a}_{0,2}(x_1^2,x_2^2).
\end{eqnarray}
Counting the degrees of these polynomials we obtain that the
division is possible only if
$$\deg(\hat{a}_{i,2-i})=4n+2-2r-2w\ge\deg(\hat{S}_{2,i}^{(n)}(x_1^2,x_2^2))=4n-2,\quad i=0,1,2$$
implying $r+w\le 2.$ Since $w>0$ is either integer or half-integer, there are only four possible cases
as follows:
\begin{enumerate}
\item $w=2,\quad r=0$;\label{c1}
\item $w=\frac{3}{2},\quad r=0$;\label{c2}
\item $w=1,\quad r=0,1$;\label{c3}
\item $w=\frac{1}{2},\quad r=0,1$.\label{c4}
\end{enumerate}
Let us  compute the degrees of quadratic terms in the system we are
considering.  Applying formula (\ref{deg}), we have
$\deg(a_{1,1}(\xi_1,\xi_2))=n-w+\frac{1}{2} $,
$\deg(a_{0,2}(\xi_1,\xi_2))=r-w $,
$\deg(a_{2,0}(\xi_1,\xi_2))=2n+1-w-r$. We know that all these
numbers should be positive integers, which enable us to determine
the possible quadratic terms.

{\bf Case \ref{c1}}. The only possible quadratic term in equation (\ref{eqs})
is $a_{2,0}$. This leads to
$$\hat{a}_{2,0}(x_1^2,x_2^2)=\hat{a}_{0,2}(x_1^2,x_2^2)=a_{2,0}(x_1^2,x_2^2),\quad
\hat{a}_{1,1}(x_1^2,x_2^2)=2a_{2,0}(x_1^2,x_2^2).$$ From formula (\ref{div}) we see that both $\hat
S_{2,i}^{(n)}(x_1^2,x_2^2)$  must divide $a_{2,0}(x_1^2,x_2^2)$. It is easy to check that
$$\gcd(\hat{S}_{2,2}^{(n)}(x_1^2,x_2^2),\hat{S}_{2,1}^{(n)}(x_1^2,x_2^2))=1.$$
So we have $\hat{S}_{2,2}^{(n)}(x_1^2,x_2^2)\hat{S}_{2,1}^{(n)}(x_1^2,x_2^2)|a_{2,0}(x_1^2,x_2^2)$. But
this is impossible since
$$\deg(a_{2,0}(x_1^2,x_2^2))=4n-2<\deg(\hat{S}_{2,2}^{(n)}(x_1^2,x_2^2)\hat{S}_{2,1}^{(n)}(x_1^2,x_2^2))=8n-4,\quad \mbox{for all $n>0$}.$$

{\bf Case \ref{c2}}. The only possible quadratic term in equation (\ref{eqs}) is $a_{1,1}$. Notice that
polynomial $\hat{S}_{2,i}^{(n)} (x_1^2,x_2^2)$, $\hat{a}_{2,0}(x_1^2,x_2^2)$ and
$\hat{a}_{0,2}(x_1^2,x_2^2)$ are symmetric and polynomial $\hat{a}_{1,1}(x_1^2,x_2^2)$ is antisymmetric
with respect to $x_1,x_2$. Due to (\ref{div}), we have
\begin{eqnarray}\label{gg}
\hat{a}_{2,0}(x_1^2,x_2^2)=-\hat{a}_{0,2}(x_1^2,x_2^2)=c_1(x_1+x_2)\hat{S}_{2,2}^{(n)}(x_1^2,x_2^2),\quad
\hat{a}_{1,1}=c_2(x_1-x_2)\hat{S}_{2,1}^{(n)}(x_1^2,x_2^2),
\end{eqnarray}
where $c_1,c_2$ are arbitrary constants. Inverting formulae
(\ref{a20h}), (\ref{a11h}) and (\ref{a02h}) or using directly
(\ref{A}) for $a_{1,1}$, we find that
\begin{equation}
\label{a11ah}
a_{1,1}(x_1^2,x_2^2)=\frac{1}{4x_2^{2n+1}}\left(2\hat{a}_{2,0}(x_1^2,x_2^2)-2\hat{a}_{0,2}(x_1^2,x_2^2)+
\hat{a}_{1,1}(x_2^2,x_1^2)-\hat{a}_{1,1}(x_1^2,x_2^2)\right),
\end{equation}
which must be a polynomial in $x_1,x_2$. Substituting (\ref{gg})
into (\ref{a11ah}), we see that this can only happen when
$c_1=c_2=0$ implying $a_{1,1}=0$. Hence our equation does not
contain quadratic terms and therefore does not possess higher
symmetries.

{\bf Case \ref{c3}}. If $r=0$, the only possible quadratic term in
equation (\ref{eqs}) is $a_{2,0}$. This leads to
\[
\hat{a}_{2,0}(x_1^2,x_2^2)=\hat{a}_{0,2}(x_1^2,x_2^2)=a_{2,0}(x_1^2,x_2^2),\quad
\hat{a}_{1,1}(x_1^2,x_2^2)=2a_{2,0}(x_1^2,x_2^2).
\]
Similar to Case \ref{c1}, we obtain that
$\hat{S}_{2,2}^{(n)}(x_1^2,x_2^2)\hat{S}_{2,1}^{(n)}(x_1^2,x_2^2)$ must divide $a_{2,0}(x_1^2,x_2^2)$,
which can only happen if
\[
\deg(a_{2,0}(x_1^2,x_2^2))=4n\ge\deg(\hat{S}_{2,2}^{(n)}(x_1^2,x_2^2)\hat{S}_{2,1}^{(n)}(x_1^2,x_2^2))
=8n-4,
\]
that is, $4n\le 4$ implying $ n=1.$  Therefore, there exists a constant $c$ such that
\[
a_{2,0}(x_1^2,x_2^2)=c\hat{S}_{2,2}^{(n)}(x_1^2,x_2^2)\hat{S}_{2,1}^{(n)}(x_1^2,x_2^2) =c\left(9
x_1^4+9 x_2^4+14x_1^2x_2^2\right).
\]
This leads to
\[
\hat{A}_{2,0}^{(j)}(x_1^2,x_2^2)=\hat{A}_{0,2}(x_1^2,x_2^2)=c(3x_1^2+3x_2^2+2x_1x_2)
\hat{N}_2^{(m_j)},\,\, \hat{A}_{1,1}^{(j)}(x_1^2,x_2^2)=c(3x_1^2+3x_2^2-2x_1x_2)\hat{N}_2^{(m_j)}.
\]
From formulae (\ref{A20}), (\ref{A11}) and (\ref{A02}) we find that
\[
A_{2,0}^{(j)}(x_1^2,x_2^2)=3c(x_1^2+x_2^2)\hat{N}_2^{(m_j)}(x_1^2,x_2^2),\qquad
A_{1,1}^{(j)}(x_1^2,x_2^2)=0 \qquad A_{0,2}^{(j)}(x_1^2,x_2^2)=2c\frac{\hat{N}_2^{(m_j)}
(x_1^2,x_2^2)}{x_1^2x_2^2}.
\]
It is easy to see that $x_1^2x_2^2$ does not divide $\hat{N}_2^{(m_j)}(x_1^2,x_2^2)$ for any integer
$m_j>1$, so $A_{0,2}(x_1^2,x_2^2)$ is not a polynomial unless $c=0$ what implies $a_{2,0}=0$. So system
(\ref{eqs}) has no quadratic terms and therefore is not integrable.

If $r=1$, then we have $a_{1,1}=0$ and therefore
\[
\hat{a}_{2,0}(x_1^2,x_2^2)=\hat{a}_{0,2}(x_1^2,x_2^2)=a_{2,0}(x_1^2,x_2^2)+a_{0,2}(x_1^2,x_2^2)(x_1x_2)^{2n+1}
\]
\[
\hat{a}_{1,1}(x_1^2,x_2^2)=2a_{2,0}(x_1^2,x_2^2)-2a_{0,2}(x_1^2,x_2^2)(x_1x_2)^{2n+1}
\]
It is easy to see that $\deg(\hat{a}_{i,2-i}(x_1^2,x_2^2))
=\deg(\hat{S}_{2,i}^{(n)}(x_1^2,x_2^2))=4n-2,\,\,i=0,1,2$. It follows from (\ref{div}) that
\begin{equation}
\label{sub}
\hat{a}_{2,0}(x_1^2,x_2^2)=\hat{a}_{0,2}(x_1^2,x_2^2)=c_1\hat{S}_{2,2}^{(n)}(x_1^2,x_2^2),\quad
\hat{a}_{1,1}(x_1^2,x_2^2)=c_2\hat{S}_{2,1}^{(n)}(x_1^2,x_2^2),
\end{equation}
where $c_1,c_2$ are constant. Inverting formulae (\ref{a20h}),
(\ref{a11h}) and (\ref{a02h}) or using directly formula (\ref{A})
for the quadratic terms, we find
\begin{eqnarray*}
&&a_{2,0}(x_1^2,x_2^2)=\frac{1}{8}\left(4\hat{a}_{2,0}(x_1^2,x_2^2)+\hat{a}_{1,1}(x_1^2,x_2^2)+
\hat{a}_{1,1}(x_2^2,x_1^2)\right),
\\
&&a_{0,2}(x_1^2,x_2^2)=\frac{1}{8x_1^{2n-1}x_2^{2n-1}}\left(4\hat{a}_{2,0}(x_1^2,x_2^2)-\hat{a}_{1,1}(x_1^2,x_2^2)-
\hat{a}_{1,1}(x_2^2,x_1^2)\right)
\end{eqnarray*}
Substituting (\ref{sub}) into the above formulae we find that $a_{0,2}(x_1^2,x_2^2)$ is a polynomial in
$x_1^2,x_2^2$ if and only if $c_2=2c_1$. Hence $ a_{0,2}(x_1^2,x_2^2)=-2c_1$. Without loss of
generality we can put $c_1=\frac{1}{2}$, so
\begin{equation}
\label{aB} a_{0,2}(x_1^2,x_2^2)=-1,\qquad
a_{2,0}(x_1^2,x_2^2)=\frac{1}{4}\sum_{j=1}^{2n}C_{2n+1}^j(x_1^2)^{j-1}(x_2^2)^{2n-j}.
\end{equation}

Therefore
\[
\hat{A}^{(j)}_{2,0}(x_1^2,x_2^2)=\hat{A}^{(j)}_{0,2}(x_1^2,x_2^2)=\frac{1}{2}(x_1^2+x_2^2)
\hat{N}_2^{(m_j)}(x_1^2,x_2^2), \,\,\,\,\,
\hat{A}^{(j)}_{1,1}(x_1^2,x_2^2)=(x_1^2+x_2^2)\hat{N}_2^{(m_j)}(x_1^2,x_2^2)
\]
and leads to
\[
A_{2,0}^{(j)}(x_1^2,x_2^2)=\frac{1}{2}(x_1^2+x_2^2)\hat{N}_2^{(m_j)}(x_1^2,x_2^2),\qquad
A_{1,1}^{(j)}(x_1^2,x_2^2)=A_{0,2}^{(j)}(x_1^2,x_2^2)=0.
\]
Thus system (\ref{eqs}) with quadratic terms (\ref{aB}) and $r=1$ possesses infinitely many type I
approximate symmetries of degree $2$ of any order $m_j>1$.

{\bf Case \ref{c4}}. The only possible quadratic term in equation (\ref{eqs}) is $a_{1,1}$. If $r=0$,
then $\deg(\hat{a}_{i,2-i}(x_1^2,x_2^2))=4n+1,\,\,i=0,1,2$. Notice that polynomials
$\hat{S}_{2,i}^{(n)} (x_1^2,x_2^2)$, $\hat{a}_{2,0}(x_1^2,x_2^2)$ and $\hat{a}_{0,2}(x_1^2,x_2^2)$ are
symmetric while polynomial $\hat{a}_{1,1}(x_1^2,x_2^2)$ is antisymmetric with respect to $x_1,x_2$. Due
to (\ref{div}), we have
\begin{eqnarray}
&&\hat{a}_{2,0}(x_1^2,x_2^2)=-\hat{a}_{0,2}(x_1^2,x_2^2)=\left(c_1(x_1^3+x_2^3)
+c_2x_1x_2(x_1+x_2)\right)\hat{S}_{2,2}^{(n)}(x_1^2,x_2^2),\label{gg1}\\
&&\hat{a}_{1,1}(x_1^2,x_2^2)=\left(d_1(x_1^3-x_2^3)+d_2x_1x_2(x_1-x_2)\right)\hat{S}_{2,1}^{(n)}(x_1^2,x_2^2),\label{gg2}
\end{eqnarray}
where $c_1,c_2,d_1,d_2$ are arbitrary constants. Substituting these
expressions into (\ref{a11ah}) as in Case \ref{c2}, we find that
$a_{1,1}(x_1^2,x_2^2)$ is a non-vanishing polynomial in even powers
of $x_1,x_2$ only when $n=1$ and
$d_1=2c_1,\,\,d_2=-\frac{4}{3}c_1,\,\,c_2=\frac{2}{3}c_1$. Without
loss of generality we can choose $c_1=\frac{3}{2}$. It follows that
\[
a_{1,1}(x_1^2,x_2)=11x_1^2+9x_2^2
\]

Therefore for $\hat{A}_{i,2-i}$ we obtain
\begin{eqnarray*}
&&\hat{A}_{2,0}^{(j)}(x_1^2,x_2^2)=-\hat{A}_{0,2}^{(j)}(x_1^2,x_2^2)=
\left(\frac{3}{2}(x_1^3+x_2^3)+x_1x_2(x_1+x_2)\right)\hat{N}_2^{(m_j)}(x_1^2,x_2^2)
\\
&&\hat{A}_{1,1}^{(j)}(x_1^2,x_2^2)=\left(3(x_1^3-x_2^3)-2x_1x_2(x_1-x_2)\right)\hat{N}_2^{(m_j)}(x_1^2,x_2^2)
\end{eqnarray*}
and hence
\[
A_{1,1}^{(j)}(x_1^2,x_2^2)=\frac{\hat{N}_2^{(m_j)}(x_1^2,x_2^2)}{x_2^2}\left(3x_2^2+2x_1^2\right),\quad
A_{2,0}^{(j)}(x_1^2,x_2^2)=A_{0,2}^{(j)}(x_1^2,x_2^2)=0.
\]
It is clear that $A_{1,1}^{(j)}(x_1^2,x_2^2)$ is not polynomial since $x_2^2$ does not divide
$\hat{N}_2^{(m_j)}(x_1^2,x_2^2)$ for any $m_j>1$. Thus equation (\ref{eqs}) does not possess an
infinite many type I approximate symmetries of degree $2$.

If $r=1$, then $\deg(\hat{a}_{i,2-i}(x_1^2,x_2^2))=4n-1,\,i=0,1,2$.
Similar as Case \ref{c2}, we obtain the formula (\ref{gg}).
Inverting (\ref{a20h}), (\ref{a11h}) and (\ref{a02h}) we find that
\begin{equation}
\label{a11ah2}
a_{1,1}(x_1^2,x_2^2)=\frac{1}{4x_2^{2n-1}}\left(2\hat{a}_{2,0}(x_1^2,x_2^2)-2\hat{a}_{0,2}(x_1^2,x_2^2)+
\hat{a}_{1,1}(x_2^2,x_1^2)-\hat{a}_{1,1}(x_1^2,x_2^2)\right),
\end{equation}
Substituting (\ref{gg}) into (\ref{a11ah2}), we find that
$a_{1,1}(x_1^2,x_2^2)$ is a non-zero polynomial in even powers of
$x_1,x_2$ if and only $c_2=2c_1$ and $n=1$. Without loss of
generality we choose $c_1=\frac{1}{2},\,c_2=1$ and hence
\begin{equation}
\label{aS} a_{1,1}(x_1^2,x_2^2)=x_1^2+3x_2^2.
\end{equation}
For $\hat{A}_{i,2-i}^{(j)}(x_1^2,x_2^2)$ we therefore obtain
\begin{eqnarray*}
&&\hat{A}_{2,0}^{(j)}(x_1^2,x_2^2)=-\hat{A}_{0,2}^{(j)}(x_1^2,x_2^2)=\frac{1}{2}(x_1+x_2)(x_1^2+x_2^2)
\hat{N}_2^{(m_j)}(x_1^2,x_2^2),
\\&&
\hat{A}_{1,1}^{(j)}(x_1^2,x_2^2)=(x_1-x_2)(x_1^2+x_2^2) \hat{N}_2^{(m_j)}(x_1^2,x_2^2)
\end{eqnarray*}
and hence
\[
A_{1,1}^{(j)}(x_1^2,x_2^2)=(x_1^2+x_2^2)\hat{N}_2^{(m_j)}(x_1^2,x_2^2),\qquad
A_{2,0}(x_1^2,x_2^2)=A_{0,2}(x_1^2,x_2^2)=0.
\]
Thus we proved that system (\ref{eqs}) with the quadratic terms $(\ref{aS})$ and $n=1,r=1$ possesses
infinitely many  type I approximate symmetries of degree $2$ of any order $m_j>1$.

Summarizing the results in the above four cases, we proved that if system (\ref{eq2}) possesses
infinitely many  type I approximate symmetries of degree $2$ of any order $m_j>1$, then it is up
re-scaling one of the systems in the following list:
\begin{eqnarray}
&&\left\{\begin{array}{l}\label{eqS3} u_{t}=v_1\\
v_t=u_2+u_1 v+3uv_1+h,\quad h=o(\ring^2)
\end{array}\right.\\
&&\left\{\begin{array}{l}\label{eqB3} u_{t}=v_1\\
v_t=u_{2n}-v^2+\frac{1}{2}\sum_{j=1}^{2n}C_{2n+1}^ju_{j-1}u_{2n-j}+h,\quad
h=o(\ring^2)
\end{array}\right.
\end{eqnarray}

Let us now assume that system (\ref{eqs}) possesses infinitely many approximate symmetries of degree
$2$ of type II of the form
\[
u_{\tau_j}=v\zeta_1^{r+m_j}+u^2A_{2,0}^{(j)}(\xi_1,\xi_2)+uvA_{1,1}^{(j)}(\xi_1,\zeta_1)+v^2A_{0,2}(\zeta_1,\zeta_2)
+o(\ring^2),\quad j=1,2,3,\ldots
\]
and $1<m_1<m_2<\cdots<m_j<m_{j+1}<\cdots$. From Theorem \ref{th1}
and Corollary \ref{Cor2}, it follows that
\[
\hat{A}^{(j)}_{2,0}(x_1^2,x_2^2)=(x_1^2+x_2^2)^r\frac{{M}_{2,2}^{(m_j)}(x_1^2,x_2^2)}{{S}_{2,2}^{(n)}(x_1^2,x_2^2)}
\hat{a}_{2,0}(x_1^2,x_2^2),\quad
\hat{A}^{(j)}_{1,1}(x_1^2,x_2^2)=(x_1^2+x_2^2)^r\frac{{M}_{2,1}^{(m_j)}(x_1^2,x_2^2)}{{S}_{2,1}^{(n)}(x_1^2,x_2^2)}
\hat{a}_{1,1}(x_1^2,x_2^2)
\]
\[
\hat{A}^{(j)}_{0,2}(x_1^2,x_2^2)=-(x_1^2+x_2^2)^r\frac{{M}_{2,2}^{(m_j)}(x_1^2,x_2^2)}{{S}_{2,2}^{(n)}(x_1^2,x_2^2)}
\hat{a}_{0,2}(x_1^2,x_2^2)
\]
must be polynomials in $x_1,x_2$ for every $m_j$. Notice that polynomials
$M_{2,i}^{(m_j)}(x_1^2,x_2^2)$ can be factorized as
\[
M_{2,i}^{(m_j)}(x_1^2,x_2^2)=x_1^2x_2^2\hat{M}_{2,i}^{(m_j)}(x_1^2,x_2^2),
\]
which leads to
\begin{eqnarray*}
&&\hat{A}^{(j)}_{2,0}(x_1^2,x_2^2)=(x_1^2+x_2^2)^r\frac{\hat{M}_{2,2}^{(m_j)}(x_1^2,x_2^2)}
{\hat{S}_{2,2}^{(n)}(x_1^2,x_2^2)} \hat{a}_{2,0}(x_1^2,x_2^2),\quad
\hat{A}^{(j)}_{1,1}(x_1^2,x_2^2)=(x_1^2+x_2^2)^r\frac{\hat{M}_{2,1}^{(m_j)}(x_1^2,x_2^2)}
{\hat{S}_{2,1}^{(n)}(x_1^2,x_2^2)} \hat{a}_{1,1}(x_1^2,x_2^2)
\\&&
\hat{A}^{(j)}_{0,2}(x_1^2,x_2^2)=-(x_1^2+x_2^2)^r\frac{\hat{M}_{2,2}^{(m_j)}(x_1^2,x_2^2)}
{\hat{S}_{2,2}^{(n)}(x_1^2,x_2^2)} \hat{a}_{0,2}(x_1^2,x_2^2)
\end{eqnarray*}
from Lemma \ref{Lem2} in Appendix A, if there exist infinitely many
$m_j$ such that the above expressions are  polynomials in $x_1,x_2$,
then $m_j=1,2,3,4,\ldots $ and
\begin{eqnarray}\label{div1}
{\hat{S}_{2,2}^{(n)}(x_1^2,x_2^2)}|\hat{a}_{2,0}(x_1^2,x_2^2),\qquad
{\hat{S}_{2,2}^{(n)}(x_1^2,x_2^2)}|\hat{a}_{0,2}(x_1^2,x_2^2)\qquad
{\hat{S}_{2,1}^{(n)}(x_1^2,x_2^2)}|\hat{a}_{1,1}(x_1^2,x_2^2).
\end{eqnarray}
We then repeat what we did for the case of type I symmetries and we obtain the similar conclusion: If
system (\ref{eqs}) possesses infinitely many type II approximate symmetries of degree $2$ of any order
$m_j>1$, then it is up re-scaling either (\ref{eqS3}) or (\ref{eqB3}).

By now we have classified all systems (\ref{eqs}) possessing infinite many approximate symmetries of
degree $2$. We know that system (\ref{eqS2}) possesses infinitely many higher symmetries of any order
of form (\ref{sym1}) and (\ref{sym2}). Notice that the quadratic  terms of (\ref{eqS2}) and  of the
system (\ref{eqS3}) coincide. By Theorem \ref{Th1}, if system (\ref{eqS3}) possesses an exact symmetry
of some order $m$ for some $h$ in its right hand side , it must coincide with the system (\ref{eqS2}),
that is, $h=-3u^2u_1$. Similarly, quadratic terms  of (\ref{Burg}) and of the system (\ref{eqB3})
coincide and system (\ref{Burg}) possesses higher symmetries of any order. Therefore, if system
(\ref{eqB3}) possesses an exact symmetry of some order for some $h$ then it coincides with
(\ref{Burg}). \hfill$\diamond$


System (\ref{Burg}) is equivalent to a linear equation $w_{tt}=\partial_x^{2n+1} w$ under the Cole-Hopf
transformation $u=(\log w)_x$. This system possesses infinitely many type I symmetries
\[
u_{\tau_m}=D_x(D_x+u)^{m-1}u,\,\,m=2,3,\ldots
\]
and type II symmetries
\[
u_{\tau_m}=D_x(v+D_t)(D_x+u)^{m-1}u.
\]
These symmetries correspond to the symmetries $w_{\tau_m}=w_m$ and
$w_{\tau_m}=w_{t\,m}$ of $w_{tt}=w_{2n+1}$.

\section{Conclusion and Discussion}
This is the first paper dealing with the global classification of the systems with two components. The
success of global classification relies on so-called implicit function theorem in \cite{sw}. The
theorem implies that the existence of infinitely many approximate symmetries of low degree together
with one symmetry leads to integrability. The proof of the theorem is straightforward and purely
algebraic. However, to check the conditions of the theorem requires to prove the irreducibility of a
family of polynomials. In this paper, Theorem \ref{Th1} can be viewed as another version of it. The
required irreducibility of polynomials $S_{s,i}^{(n)}(x_1^2,\ldots,x_s^2)$ defined by (\ref{S}) is
proved by algebraic geometrical arguments in Proposition \ref{Pro4}.

The formal diagonalisation approach proposed in this paper can be
applied to large classes of systems with  the diagonalisable linear
part. This type of equations is generic and it has not been treated
by other classification methods. Our approach enables us to
explicitly write down the necessary conditions for integrability. We
are able to prove that these conditions are indeed sufficient.
Resolution of these conditions leads us to the ultimate
classification result.

We have not applied this approach to the systems with nilpotent
linear part yet. However, we do not expect any difficulties in this
case. The final goal will be to combine these two classes and to
obtain the global classification of homogeneous polynomial
evolutionary systems of two components as it has been done for the
scalar evolutionary equations \cite{sw}.

\section*{Appendix A}
Here we prove the lemmas required in the proof of Theorem \ref{Th2}.
which is based on the theorem of Lech-Mahler and
its straightforward corollary.
\begin{The}[Lech, Mahler]\label{lechmahler}
Let $A_1,A_2,\ldots,A_n\in\mathbb{C}$ and $a_1,a_2,\ldots,a_n$ be
non-zero complex numbers. Suppose that none of the ratios $A_i/A_j$
with $i\ne j$ is a root of unity. Then the equation
$$a_1A_1^m+a_2A_2^m+\cdots+a_nA_n^m=0$$ in the unknown integer
$m$ has finitely many solutions.
\end{The}
\begin{Cor}\label{abc}
Let $A,B,C\in\mathbb{C}$ and $a,b,c$ be non-zero complex numbers.
Suppose that the equation $$a A^m+b B^m+c C^m=0$$ has infinitely
many integers $m$ as solution. Then $A/B$ and $A/C$ are both roots
of unity.
\end{Cor}

We simplify our notations by writing $S^+=S_{2,2}^{(n)}$, $S^-=S_{2,1}^{(n)}$,
$M^{+}_{m_j}=M_{2,2}^{(m_j)}$ and $M^{-}_{m_j}=M_{2,1}^{(m_j)}$.
\begin{Lem}\label{Lem1}
Consider an infinite sequence of polynomials
\begin{equation}
\label{Mseq} N_2^{(m_j)}(x_1^2,x_2^2)=(x_1^2+x_2^2)^{m_j}-x_1^{2m_j}-x_2^{2m_j}, \quad
1<m_1<m_2<\cdots<m_j<m_{j+1}<\cdots
\end{equation}
If there exist infinite many $m_j$ such that each $N_2^{(m_j)} (x_1^2,x_2^2)$ has nontrivial common
divisor with the polynomial $S^{\pm}(x_1^2,x_2^2)$ defined by (\ref{S}), Then $m_j=2,3,4, \ldots$ and
$\gcd \{S^{\pm}(x_1^2,x_2^2), N_2^{(2)}(x_1^2,x_2^2),N_2^{(3)}(x_1^2,x_2^2),\ldots \}=x_1^2x_2^2$.
\end{Lem}
{\bf Proof}.  Since the polynomials $S^{\pm}(x_1^2,x_2^2)$ and $N_2^{(m_j)}(x_1^2,x_2^2)$ are
homogeneous polynomials in $x_1,x_2$ it is convenient to consider them in affine coordinate
$\frac{x_1}{x_2}$. If they have nontrivial common divisor, these exist $q$ as a common root, that is,
\begin{eqnarray*}
&&S^{\pm}(q)=(1+q^2)^{2n+1}-(1\pm q^{2n+1})^2=0\\
&&N_2^{(m_j)}(q)=(1+q^2)^{m_j}-1-q^{2m_j}=0,
\end{eqnarray*}
which is required to be satisfied for infinitely many integers
$m_j>1$. Applying the Lech-Mahler theorem to it we find
that this is possible if one of the listed cases holds:
\begin{itemize}
\item $q=0$ (double root), $m_j=2,3,4,\ldots$,
\item $q=\pm i$, $m_j=1 \mod 2 $,
\item $q=e^{\frac{\pi i}{3}},\,e^{\frac{2\pi i}{3}},\,e^{\frac{4\pi i}{3}},\,e^{\frac{5\pi
i}{3}}$, $m_j=5 \mod 6$,
\item $q=e^{\frac{\pi i}{3}},\,e^{\frac{2\pi i}{3}},\,e^{\frac{4\pi i}{3}},\,e^{\frac{5\pi
i}{3}}$ (double roots), $m_j=1\mod 6$.
\end{itemize}
The last three sets of roots do not satisfy the equation $S^{\pm}(q)=0$ for any $n>0$, while $q=0$ is a
double roots of $S^{\pm}(q)$. Lemma is proved. $\diamond$

\begin{Lem}\label{Lem2}
Consider an infinite sequence of polynomials
\begin{eqnarray}
\label{M1seq}
&&M^{\pm}_{m_j}(x_1^2,x_2^2)=(x_1^2+x_2^2)(x_1^{2n+1}\pm
x_2^{2n+1})-x_1^{2n+2m+1}-(\pm x_2^{2n+2m+1}),
\\ \nonumber &&\quad\quad\quad
0<m_1<m_2<\cdots<m_j<m_{j+1}<\cdots
\end{eqnarray}
If there exist infinite many $m_j$ such that each
$M^{+}_{m_j}(x_1^2,x_2^2)$ (or $M^{-}_{m_j}(x_1^2,x_2^2)$)
has nontrivial common divisor with the polynomial
$S^{+}(x_1^2,x_2^2)$ (or $S^{-}(x_1^2,x_2^2)$ )
defined by (\ref{S}),
Then $m_j=1,2,3,4, \ldots$ and
\begin{eqnarray*}
&&\gcd \{S^{+}(x_1^2,x_2^2),
M^{+}_1(x_1^2,x_2^2),M^{+}_2(x_1^2,x_2^2),\ldots \}=x_1^2x_2^2;\\
&&\gcd \{S^{-}(x_1^2,x_2^2),
M^{-}_1(x_1^2,x_2^2),M^{-}_2(x_1^2,x_2^2),\ldots \}=x_1^2x_2^2.
\end{eqnarray*}
\end{Lem}
{\bf Proof}. Similar as in the proof of Lemma \ref{Lem1}, we consider
polynomials $S^{\pm}(x_1^2,x_2^2)$ and $M_{m_j}^{\pm}(x_1^2,x_2^2)$
in affine coordinate $\frac{x_1}{x_2}$.
If  $S^{+}(x_1^2,x_2^2)$ and $M_{m_j}^{+}(x_1^2,x_2^2)$ have nontrivial common
divisor, these exist $q$ as a common root, that is,
\begin{eqnarray*}
&&S^{+}(q)=(1+q^2)^{2n+1}-1-q^{4n+2}-2q^{2n+1}=0\\
&&M^{+}_{m_j}(q)=(1+q^2)^{m_j}(1+q^{2n+1})-(q^2)^{m_j} q^{2n+1}-1=0,
\end{eqnarray*}
which is required to hold for infinitely many positive integers
$m_j$. We apply the Lech-Mahler theorem to it and obtain:
\begin{itemize}
\item $q=0$ (double root), $m_j\in\bbbz_{+}$,
\item $q^{2n+1}=-1,\,q^{2m_j}=1$,
\item $q=e^{\frac{\pi i}{3}},e^{\frac{5\pi i}{3}}$ and $n=1\mod 3,\,\,\,m_j=0\mod
3$ or $n=2\mod 3,\,\,\, m_j=0\mod 6,\,1\mod 6$ or $n=0\mod 3,\,m_j=0\mod
6,\,5\mod 6$,
\item $q=e^{\frac{2\pi i}{3}},e^{\frac{4\pi i}{3}}$ and $n=1\mod 3,\,\,\,m_j=0\mod
6$ or $n=2\mod 3,\,\,\, m_j=0\mod 6,\,4\mod 6$ or $n=0\mod 3,\,m_j=0\mod
6,\,2\mod 6$.
\end{itemize}
It is easy to check that the roots from the last three cases are not
the roots of $S^+(q)$ for any integer $n\ge 1$, while $q=0$ is a
double root of $S^+(q)$. Therefore, we have $m_j=1,2,3, \ldots$ and
$\gcd \{S^{+}(x_1^2,x_2^2),
M^{+}_1(x_1^2,x_2^2),M^{+}_2(x_1^2,x_2^2),\ldots \}=x_1^2x_2^2$.
Similarly, we can prove the second half of the lemma. $\diamond$

\begin{center}
{\large UNIVERSITY OF KENT}
\end{center}

\end{document}